\documentclass[11pt]{article}
\usepackage{epsfig}
\usepackage{graphicx}
\usepackage{wrapfig}
\usepackage{amsmath}
\usepackage{subfigure}
\usepackage{bm}

\def\la{\langle}
\def\ra{\rangle}
\def\be{\begin{equation}}\def\ba{\begin{eqnarray}}
\def\ee{\end{equation}}\def\ea{\end{eqnarray}}
\def\ben{\begin{enumerate}}\def\bitem{\begin{itemize}}
\def\een{\end{enumerate}}\def\eitem{\end{itemize}}
\def\no{\nonumber\\}

\def\calL{{\cal L}}

\newcommand{\e}{{\mbox{e}}}

\def\roughly#1{\mathrel{\raise.3ex\hbox{$#1$\kern-.75em%
\lower1ex\hbox{$\sim$}}}}

\def\la{\langle}\def\ra{\rangle}
\def\deriv{\partial}

\def\exp{\mbox{e}}

\def\be{\begin{equation}}
\def\ee{\end{equation}}
\def\bea{\begin{eqnarray}}
\def\eea{\end{eqnarray}}

\def\sF{{{\rm F}\!\!\!\!\hskip.8pt\hbox{\raise1pt\hbox{/}}\,}}

\def\a{\alpha}

\def\e{\epsilon}

\def\o{\omega}

\def\r{\rho}

\def\t{\tau}

\def\O{\Omega}

\begin{document}
\begin{titlepage}
\begin{center}

 \vskip 1.5cm

{\Large \bf The Dropping of In-Medium Hadron Mass in Holographic QCD}
\vskip 1. cm
  { Kwanghyun Jo$^{a}$, Mannque Rho$^{a,b}$, Yunseok Seo$^{c}$ and Sang-Jin Sin$^{a}$}

\vskip 0.5cm

(a){\it  Department of Physics, Hanyang University, Seoul 133-791, Korea,}\\
(b){\it Institut de Physique Th¢¥eorique, CEA Saclay, F-91191 GIf-sur-Yvette, France}\\
(c){\it  Center for Quantum Spacetime (CQUeST), Sogang University, Seoul 127-742, Korea}

\end{center}

\centerline{(\today) }
\vskip 1cm
\vspace{1.0cm plus 0.5cm minus 0.5cm}

\begin{abstract}
We study the baryon density dependence of the vector meson spectrum using the D4/D6 system together with the compact D4 baryon vertex. We find that the vector meson mass decreases almost linearly in density at low density for small quark mass, but saturates to a finite non-zero value for large density.
We also compute the density dependence of the $\eta\prime$ mass and the $\eta\prime$ velocity. We find that in medium, our model is consistent with the GMOR relation up to a few times the normal nuclear density. We compare our hQCD predictions with predictions made based on hidden local gauge theory that is constructed to model QCD.
\end{abstract}

\end{titlepage}

\newpage
\section{Introduction}\label{intr}
If light-quark hadron masses are dynamically generated by the spontaneous breaking of chiral symmetry as it is commonly accepted in QCD, then it should be possible to ``unbreak" the symmetry by dialing such external conditions as temperature and/or baryonic density. This means that the mass should change as the quark condensate -- that is the order parameter of chiral symmetry -- changes when the the external background such as temperature and/or baryonic density is dialled. This phenomenon that reflects the vacuum change in strong-interaction physics is one of the main lines of research in nuclear and particle physics.

In the case of temperature, the great progress in lattice QCD calculations together with recent experiments at relativistic heavy-ion colliders has enabled one to start understanding what is happening at high temperature. However the situation is totally different for dense baryonic matter such as one expects to encounter in compact stars. Because of the famous sign problem, lattice methods cannot access cold baryonic matter at high density. While there are a large number of models anchored on phenomenological or effective field theory approaches available in the literature,  there are no known reliable tools to address the density that exceeds the normal nuclear matter density $n_0\simeq 0.16$ fm$^{-3}$.
Thus the density regime involved in compact stars is theoretically a wide-open area. On the experimental side, the forthcoming accelerators (such as FAIR/GSI, J-PARC etc.) will probe higher densities, but it will be sometime before results will become available to help guide the theoretical efforts.

It is the objective of this paper to approach the hadron-mass problem by a holographic dual method along the line of recent developments in
using gravity/gauge duality  in top down approach \cite{Kruczenski:2003uq,SS}. The advantage of the holographic approach is that it can handle -- albeit in certain limits -- the strong-coupling aspects of gauge theory relevant for dense matter in terms of weak coupling bulk gravity solutions. The logical procedure would then be the following. First one checks that the hQCD model so constructed correctly postdicts the observables probed by nuclear physics experiments, which means up to the normal nuclear matter density $n_0$. It turns out that this is not an easy task as we will see. Once the first step is taken, then one could proceed to make predictions beyond the normal density $n_0$ in as yet un-probed regime.  The dense system we shall holographically realize is the confining D4 with the compact baryonic D4 brane geometry discussed in \cite{Seo:2008qc, Seo:2009kg}.

The reason for resorting to the D4D6 model is because it can accommodate the quark mass geometrically
and we can discuss the  condensate $\la{\bar\psi}\psi\ra$, that we call ``chiral condensate''(CC).
Note, however, that since the quark in this model is not a Weyl fermion at zero quark mass limit, chirality is not defined.
Thus it is perhaps more proper to call it ``scalar density" of fermion rather than chiral condensate given that
the symmetry involved is not the axial symmetry but rahter the rotational symmetry of the brane embedding.
Nevertheless one may think of it in terms of chiral symmetry \cite{Kruczenski:2003uq,evans}, because, based on it, one can still obtain the Gell-Mann-Oakes-Renner relation, which is one of the signposts of chiral symmetry in hadron physics.

The plan of this paper is as follows. In Section 2, we describe what has been predicted in the boundary (gauge) sector and where the theory stands vis-\`a-vis with confrontation with nature. This will be done anchored on an effective field theory of strong interactions that is considered to at least partially represent QCD. It will be referred to as ``BR scaling" standing for a circle of related ideas addressing the issue of chiral symmetry in dense medium. The setup of brane embeddings figuring in our holographic approach to dense matter is given in Section 3. In Section 4, we apply the formalism to in-medium spectra of the pseudoscalar and transverse vector mesons that we are concerned with. Numerical analysis of the mass vs. density relation is made in Section 5. In Sections 6 and 7, we calculate the pion velocity and dispersion relation. Comparison with the predictions made by the class of approaches in line with Brown-Rho scaling is given in Section 8 with concluding remarks. Several details left out in the main text are relegated to Appendix.

\section{In-Medium Scaling In ``QCD-Motivated" Approach}\label{BRS}
In order to streamline our line of reasoning and specify what we are going to compare with, we first describe the presently available approach to the in-medium mass problem from a ``QCD-motivated" angle\footnote{This is a bit misleading terminology if not a misnomer in that the circle of reasoning involved is not directly backed by QCD-proper calculations. It is based on various yet-to-be confirmed results inferred from effective field theories that are supposed to model QCD. We shall adhere to that terminology to distinguish it from hQCD approaches.}. As mentioned above, there are no model-independent gauge theory (QCD) tools for dense matter (except at the extreme -- perhaps physically irrelevant -- density regime where perturbative QCD is applicable). What we will compare with is a scaling relation that combines different lines of reasoning anchored on the notion of hidden local symmetry (HLS for short)~\cite{HY:PR} combined with chiral perturbation. Specifically the approach we will focus on is what we shall refer to as ``Brown-Rho scaling approach" (BRSA for short)~\cite{BR91}.

Originally, the BRSA to how hadron masses behave as the vacuum structure changes at increasing density was derived from the skyrmion model implemented with a scalar meson, dilaton, associated with the QCD trace anomaly~\cite{BR91}. It can be given a simpler interpretation in terms of the constituent quark model to which the skyrmion description should be equivalent in the large $N_c$ limit in matter-free space and presumably in medium. In the modern version based on large $N_c$ considerations~\cite{weinberg}, the constituent quark model should work not only for very low energy but also for intermediate energy in which deconfinement and spontaneously broken chiral symmetry co-exist. In the large $N_c$ limit, the constituent quark mass $m_Q$ and the constituent  axial-vector coupling constant $g_A$ are independent of $N_c$, and baryons (mesons) are loosely bound three-quark (two-quark) states. In \cite{weinberg}, the $N_c$-independent $m_Q$ is a given quantity in the matter-free vacuum, and there is nothing one can say from the analysis how $m_Q$ will behave in dense medium. The assumption we make here is that being dynamically generated by chiral symmetry breaking, $m_Q$ will vanish when chiral symmetry is ``unbroken," i.e., when the quark condensation $\la\bar{q}q\ra\rightarrow 0$. Suppose further that the condensate can be driven to drop as density is increased, going to zero (in the chiral limit) at a critical density $n_c$. Thus $m_Q$ will continue to drop as density increases.

Now the constituent quark is a quasiparticle in the sense of many-body theory. If one applies the quasiparicle picture to $AN_c$ quark systems where $A>1$ for nuclei and nuclear matter, the issue of how hadron masses behave in $A>1$ systems would crucially depend on how the quasiparticle mass behaves in medium. Thus in the naivest scheme where the quasiparticle picture is assumed to apply not just to a single hadron but also to many-nucleon systems, the ratio of the masses in medium (with density $n$) denoted with the asterisk should scale
\bea
m_M^* (n)/m_M\approx m_B^* (n)/m_B\equiv \Phi (n)\label{BR}
\eea
with the number of constituent quarks involved in the hadrons dropping out.
 Here the subscript $M$ denotes light-quark mesons other than the pion and $B$ light-quark baryons, i.e., nucleons. This is identical to the scaling proposed by Brown and Rho a long time ago~\cite{BR91} based on a dilaton-implemented skyrmion model. Even if the constituent quark mass is independent of $N_c$ in the vacuum, it is not known how it behaves in $N_c$ in medium and how it depends on the quark condensate $\la\bar{q}q\ra$ as the condensate goes toward zero in the chiral limit as the chiral restoration is approached.
 Typically $\Phi(n)=1-Cn/n_0 $ with  $C= 0.1 \sim 0.3$ \cite{Holt:2006ii}.
 One can offer a plausible argument that the scaling factor $\Phi(n)$ at low density is related to the pion decay constant in medium
\be
\Phi(n)_{BR}\approx f_\pi^* (n)/f_\pi. \label{BR1}
\ee
However it is not really known what happens at a density greater than that of normal nuclear matter, except near the critical density for chiral phase transition for which hidden local symmetry theory~\cite{HY:PR} makes a simple prediction. In  this (HLS) model -- which provides a field theory support to BR scaling, local gauge symmetry
plays an important role in giving the scaling
\bea
m_\rho^*(n)/m_\rho\sim \la\bar{q}q\ra^*(n)/\la\bar{q}q\ra\ \ \ {\rm as} \ \ n\rightarrow n_c.\label{VM}
\eea
This would imply that the vector meson mass goes to zero at the chiral transition. To distinguish this from the BRS prediction, we shall call this ``VM" prediction.\footnote{VM stands for the vector manifestation fixed point in hidden local symmetry theory~\cite{HY:PR}.}
One of the motivations of this work is to test the relations (\ref{BR}), (\ref{BR1}) and (\ref{VM}) using holographic QCD.

In comparing the hQCD prediction that we are making with the BR/VM prediction (\ref{BR})-(\ref{VM}), we should clarify how the scaling behaviors could be assessed. In order to be able to say which scaling is correct and which is not, one should know how the scalings appear in physical observables. For instance, in the BRS approach, the scaling figures in nuclear physics as parameters of the effective Lagrangian used in nuclear forces and in correlation functions. Only when the quasiparticle picture is a very good approximation will the scaling appear in a simple transparent form. Otherwise it will be compounded with complex many-body dynamics and it will not be an easy task to single out the ``basic" quantity that is addressed here. With respect to the present situation in confronting nature, one can say that the scaling (\ref{BR}) is consistent with -- though not confirmed by -- a variety of nuclear observables but there is nothing yet to support or refute the scaling (\ref{VM}). To address this scaling in hQCD is the purpose of this paper. Ultimately it will be up to experiments to establish the correct scaling relation and confirm or falsify the predictions of our model.

\section{Computational setup: D4 and D6 and the Baryon vertex}\label{setup}
In this section, we detail our strategy in addressing dense matter. We will focus on the low temperature, highly dense system, holographically realized as the confining D4 with the compact baryonic D4 brane geometry \cite{Seo:2008qc, Seo:2009kg}. The non-supersymmetric, confining D4 brane background with Euclidean signature, is given by
\begin{eqnarray}
ds^{2}
&=&\left(\frac{U }{R }\right)^{3/2}\left(\eta_{\mu\nu}dx^\mu dx^\nu + f(U) dx_4^{2} \right)
+\left(\frac{R}{U }\right)^{3/2}\left( \frac{dU^2}{ f(U)} +U^2 d\Omega_4^2\right) \cr
e^\phi&=&g_s\left(\frac{U }{R }\right)^{3/4},\quad F_4 =\frac{2\pi N_c}{\Omega_4}\epsilon_4, \;\; f(U)=
1-\Big(\frac{U_{KK}}{U}\Big)^{3}, \;\; R^3=\pi g_s N_c l_s^3.
\label{adsm}
\end{eqnarray}
This background is related to the geometry with Lorentzian signature by the double Wick rotation
\be
x_4 \longleftrightarrow t,~~~~~~~~t \longleftrightarrow x_4,~~~~~~~~~U_0 \longleftrightarrow U_{KK}.
\ee
The Kaluza-Klein mass is defined as the compactified radius of the $x_4$ direction:
$M_{KK}=\frac{3}{2}\frac{U^{1/2}_{KK}}{R^{3/2}}$.
And the bulk parameters ($U_{KK}, g_s, R$) are related to the gauge theory parameters
($M_{KK}$, $g_{YM}$, $\lambda$) as
\be
g_s=\frac{\lambda}{2\pi l_sN_c M_{KK}}, \quad U_{KK}=\frac{2}{9}\lambda M_{KK} l_s^2,
\quad R^3=\frac{\lambda l_s^2}{2M_{KK}} ,\quad \lambda=g_{YM}^{2}N_{c}.
\ee
By introducing a dimensionless coordinate $\xi$ as $\frac{d U^2}{f(U)} = \left(\frac{U}{\xi}\right)^2 d\xi^2$, the bulk geometry is rewritten
\be\label{d4bgmetric}
ds^2 = \left(\frac{U }{R }\right)^{3/2}\left(dt^2 +d\vec{x}^2 + f(U) dx_4^{2} \right)
+\left(\frac{R}{U }\right)^{3/2}\left(\frac{U}{\xi}\right)^2\left(d\xi^2 +\xi^2 d\Omega_4^2\right),
\ee
and $U$, $\xi$ are related
\ba
\left(\frac{U}{U_{KK}}\right)^{3/2} &=& \frac{1}{2}\left(\left(\frac{\xi}{\xi_0}\right)^{3/2}+\left(\frac{\xi_0}{\xi}\right)^{3/2}\right) , \quad f = \left(\frac{1-(\xi/\xi_0)^{-3}}{1+(\xi/\xi_0)^{-3}}\right)^2 = \frac{\omega_-^2}{\omega_+^2}\no
U &=& \frac{U_{KK}}{4^{1/3}}\frac{\xi}{\xi_0} \left(1+\left(\frac{\xi_0}{\xi}\right)^3\right)^{2/3} =  \xi ~ \omega_+^{2/3} , \quad \xi_0 = \frac{U_{KK}}{4^{1/3}}.
\ea

\begin{table}[h]
\begin{center}
\begin{tabular}{|c|c|c|c|c|c|c|c|c|c|c|}
\hline
&\multicolumn{4}{|c|}{Boundary} & S$^1$ & r(S$^4$)&\multicolumn{4}{|c|}{S$^4$}\\
\hline
coordinate & x$^0$ & x$^1$ & x$^2$ & x$^3$ & x$^4$ & U ($\sim \xi$) & $\theta$ & $\psi_1$ & $\psi_2$ & $\psi_3$ \\
\hline
Backgr D4 & $\bullet$ & $\bullet$ & $\bullet$ & $\bullet$ & $\bullet$  &&&&& \\
\hline
Baryonic D4 & $\bullet$ &&&&&& $\bullet$ & $\bullet$ & $\bullet$ & $\bullet$ \\
\hline \hline
&\multicolumn{4}{|c|}{Boundary} & $S^1$ &\multicolumn{3}{|c|}{$R^3$} & \multicolumn{2}{|c|}{$R^2$} \\
\hline
coordinate & x$^0$ & x$^1$ & x$^2$ & x$^3$ & x$^4$ & $\rho$ & $\theta_1$ & $\theta_2$ & y & $\phi$ \\
\hline
Flavor D6 &$\bullet$&$\bullet$&$\bullet$&$\bullet$&&$\bullet$&$\bullet$&$\bullet$&& \\
\hline
\end{tabular}
\end{center}\caption{The brane profile : the background D4, the compact D4 and the probe D6 \label{braneprofile1}}
\end{table}

\subsection{The compact D4 baryon vertex }
In confining phase, the baryon vertex is a D4 brane wrapping on $S^4$, see Table \ref{braneprofile1}.
For the computational convenience, we expand the internal $S^4$ as $R^1 \times S^3$. The background metric (\ref{d4bgmetric}) takes the form
\be
ds^2 = \left(\frac{U}{R}\right)^{3/2}\left(dt^2 +f dx_4^2 +d\vec{x}^2  \right)
+R^{3/2}\sqrt{U}\left(\frac{d\xi^2}{\xi^2} +d\theta^2 +\sin^2\theta d\Omega_{3}^{2}\right),
\ee
The induced metric on the compact D4 brane is
\be\label{d4met}
ds_{D4}^2 = \left(\frac{U}{R}\right)^{3/2}dt^2
+R^{3/2}\sqrt{U}\left[\left(1+\frac{\xi'^2}{\xi^2}\right)d\theta^2 +\sin^2\theta d\Omega_{3}^{2}\right],
\ee
and $\xi' =\partial \xi/\partial \theta$. The DBI action for a single D4 brane on which $N_c$ fundamental string is attached is
\ba \label{bary-d4}
S_{D4} &=& -\mu_4 \int e^{-\phi} \sqrt{{\rm det}(g+2\pi \alpha' F)}+\mu_4 \int  A_{(1)}\wedge G_{(4)} \cr\cr
&=& \tau_4 \int dt d\theta \sin^3\theta
\left[-\sqrt{ \o_+^{4/3} (\xi^2 +\xi'^2)-\tilde{F}^2}
+3 \tilde{A}_t \right] = \int dt {\cal L}_{D4},
\ea
where
\ba
\tau_4 &=& \mu_4 \O_3 g_s^{-1} R^{3}\frac{U_{KK}}{2^{2/3}}=\frac{N_c U_{KK}}{2^{8/3}(2\pi l_s^2)}, \quad \quad \mu_4 = \frac{1}{(2\pi)^4l_s^5} \cr\cr
\tilde{A}_t &=& \frac{ 2\pi\a'}{\xi_0} A_t \qquad
\tilde{F} = \frac{2\pi \alpha'}{\xi_0} F_{t\theta}.
\ea
Since it reduces the number of equations to be solved, we introduce the Legendre transformed `Hamiltonian'
\ba\label{d4h}
{\cal H}_{D4} &=&\tilde{F}\frac{\partial {\cal L}_{D4}}{\partial \tilde{F}}-{\cal L}_{D4}\cr\cr
&=&\t_4 \int d\theta \sqrt{\omega_+^{4/3} (\xi^2 +\xi'^2)}\sqrt{D(\theta)^2+ \sin^6\theta},
\ea
where the dimensionless displacement, D($\theta$), is defined by
\be
\frac{\partial {\cal L}_{D4}}{\partial \tilde{F}} \equiv -D(\theta) = \frac{\sin^3\theta \tilde{F}}{\sqrt{({\o_-^2}/{\o_+^{2/3}})(\xi^2 +\xi'^2)
-\tilde{F}^2}} .
\ee
Then the equation of motion for the gauge field is
\be
\partial_{\theta}D(\theta)=-3\sin^3 \theta,
\ee
from which by integrating, we get the solution D($\theta$)
\be\label{displacement}
D(\theta)=2(2\nu -1)+3(\cos\theta -\frac{1}{3}\cos^3\theta),
\ee
where the integration constant $\nu$ determines the number of fundamental strings. Note that $\nu N_c$ strings are attached at the south pole ($\theta=0$) and $(1-\nu)N_c$ strings at the north pole ($\theta=\pi$). Here we set $\nu =0$ and then impose smooth boundary
conditions ($\xi'(0)=0$ and $\xi(0)=\xi_0$) at the south pole, which means there is no string at the south pole or equivalently all fundamental strings are attached at the north pole. Note that the numerical solutions of baryon D4 are parameterized by the initial value, $\xi_0$.

We considered a single compact D4 brane with $N_c$ fundamental strings attached at the north pole. The tensions of D4 and fundamental strings are balanced to have a stable configuration. Assume that all fundamental strings are dangled at the same point\footnote{If we take a D4-D6 system with a confining background and consider the baryon vertex as compact D4, both D4 and D6 get deformed due to the interaction between them through the fundamental string (F1). When the strings and D-branes are in contact, the length of the F1 tends to be zero while the D-branes are deformed to replace them for the minimum energy configuration. Therefore the strings connecting D4 and D6 have zero length and D4 and D6 are in contact at a point.}, a cusp at the north pole of D4, and the location of the cusp is $U_c$. The force from the D4 brane tension can be obtained by varying the Hamiltonian of the D4 brane with respect to $U_c$ while keeping other variables constant;
\be\label{force-d4}
F_{D4} = \frac{\partial{{\cal H}}}{\partial U_c} \Bigg|_{\mbox{fix~other~values}} = N_c T_F \left(\frac{1+\xi_c^{-3}}{1-\xi_c^{-3}}\right)
\frac{\xi_c'}{\sqrt{\xi_c^2 +\xi_c'^2}},
\ee
where $ T_F =\frac{2^{2/3}\t_4 }{N_c U_{KK}}$ is the tension of the fundamental strings. To have a stable configuration, the system needs something to support the compact D4 because the tension of fundamental string is always greater or equal to the D4 brane tension. That something is probe D6 brane.

\subsection{Probe D6 brane}
Now, we put the probe D6 brane where the other endpoints of fundamental strings are
attached. The string endpoints can be understood as point charges on D6 brane with $A_{t}$ being the gauge potential that couples to this point source. Then the bulk metric (\ref{d4bgmetric}) can be written as
\be
ds^2 = \left(\frac{U }{R }\right)^{3/2}\left(dt^2 +d\vec{x}^2 + f(U) dx_4^{2} \right)
+\left(\frac{R}{U }\right)^{3/2}\left(\frac{U}{\xi}\right)^2\left(d\r^2 +\r^2 d\Omega_2^2+dy^2 +y^2 d\phi^2\right),
\ee
where the D6 brane world volume coordinates are $(t,\vec{x},\r,\theta_{\a})$. We assume that the only function that depends on $\rho$ is $y(\rho$), with the transverse direction $\phi$ set to zero. The induced metric on D6 brane is
\be \label{D4sol1}
ds^2_{D6} =\left(\frac{U}{R}\right)^{3/2}(dt^2 + d\vec{x}^2) +\left(\frac{R}{U}\right)^{3/2}\left(\frac{U}{\xi}\right)^2
\left[(1+\dot{y}^2)d\r^2 +\r^2 d\O_2^2\right],
\ee
where
\be \label{D4parameters1}
\dot{y}=\partial y/\partial\r , \quad R^3 = \frac{1}{2} \frac{\lambda l_s^2}{M_{KK}}, \quad g_s = \frac{1}{2\pi} \frac{\lambda}{N_c M_{KK}l_s}, \quad U_{KK}=\frac{2}{9} \lambda M_{KK} l_s^2.
\ee
The DBI action for $N_f$ D6 brane is
\ba \label{D6DBIaction1}
S_{D6}=  \int dt {\cal L}_{D6} &=& -N_f \mu_6 \int e^{-\phi}\sqrt{{\rm det}(g+2\pi \a' F)} \cr\cr
&=& -\t_6 \int dtd\r \r^2  \o_+^{4/3} \sqrt{\o_+^{4/3} (1+\dot{y}^2)-\tilde{F}^2},
\ea
where
\ba \label{d6tension}
\t_6 =N_f \mu_6 V_3  \O_2 g_s^{-1} \xi_0^3 ,\quad \mu_6 = \frac{1}{(2\pi)^6 l_s^7}, \quad
\tilde{F}=\frac{2\pi\a'F_{t\r}}{\xi_0}.
\ea
To see more clearly the dimensionality of the Lagrangian and other quantities, we introduce the dimensionless coordinates ($\tilde{\xi},\tilde{U},\tilde{\rho},\tilde{y}$) rescaled by $\xi_0, U_{KK}$,
\be
U = U_{KK} \tilde{U}, \quad \xi = \xi_0 \tilde{\xi},\quad y = \xi_0 \tilde{y}, \quad \rho = \xi_0 \tilde{\rho}.
\ee
Hereafter we shall drop the tilde in all dimensionless variables unless ambiguous.
We define the constant of motion Q and its dimensionless partner $\tilde{Q}$ from the equation of motion for $\tilde{F}$ :
\be \label{qtildedefinition}
\frac{\partial \calL_{D6}}{\partial \tilde{F}}
=\frac{ \r^2 \o_+^{4/3}\tilde{F}}{\sqrt{\o_+^{4/3} (1+\dot{y}^2)-\tilde{F}^2}}
\equiv \tilde{Q}.
\ee

This Q represents the number of point sources (number of fundamental strings) and is related to
$\tilde{Q}$ by
\be \label{QtildeQrelation}
\tilde{Q}=\frac{\xi_0 Q}{2\pi\a'\t_6} = \frac{\lambda M_{KK}}{4^{1/3}9\pi ~\tau_6} Q.
\ee
To compare our model with the real world, we identify the number of fundamental strings Q with the number of quarks. Then the baryon density $n_B$ is expressed in terms of Q
\be
n_B = \frac{N_B}{V_3} = \frac{Q}{N_C V_3} = \frac{32^{1/3}}{81 (2\pi)^3} \lambda M_{KK}^3
\tilde{Q} = 0.138 ~\lambda~ \left(\frac{M_{KK}}{1.04 ~\mbox{GeV}}\right)^3 ~  \tilde{Q} ~ n_0,
\ee
with the normal nuclear density given by $n_0$ = 0.16 fm$^{-3}$ = 1.28 $\times$ 10$^{-3}$ GeV$^3$. We note that from \cite{Kruczenski:2003uq}, the validity of the solution Eq.(\ref{D4sol1}) is guaranteed if the following condition is satisfied,
\be
\frac{\lambda}{N_c^2} \ll \frac{1}{\lambda} \ll 1.
\ee
And this condition is rewritten as
\ba \label{validitycondit}
\frac{\lambda}{N_c^2} \ll \frac{1}{\lambda} & \longleftrightarrow & (9U_{KK}M_{KK})^{1/3}\left(\frac{R}{l_s}\right)^2 \gg 1, \no
\frac{1}{\lambda} \ll 1 & \longleftrightarrow & \frac{N_c}{2^{1/4}} \gg \lambda
\ea
where we have used Eq.~(\ref{D4parameters1}). It means $\lambda$ should be larger than 1 but less than $N_c$. Therefore if we take $N_c$=3, the 't Hooft coupling $\lambda$ cannot be greater than 2.5 for the validity of the solution.

The ``Hamiltonian"\footnote{This is not the energy of the system but a kind of on-shell action. Instead of solving the equation of motion for $A_t$, we use Eq. (\ref{qtildedefinition}), the constant of motion, which is the solution of $A_t$.} can be obtained by Legendre transformation;
\ba
{\cal H}_{D6} &=&\tilde{F}\frac{\partial S_{D6}}{\partial \tilde{F}}-S_{D6} \cr\cr
&=& \t_6 \int d\r \sqrt{\o_+^{4/3}\left(\tilde{Q}^2+\r^4 \o_+^{8/3}\right)}\sqrt{1+\dot{y}^2} \cr\cr
&=& \t_6 \int d\r V(\r)\sqrt{1+\dot{y}^2}. \label{effectiveTension}
\ea
The equation of motion for the embedding y($\rho$) is
\be\label{emod6}
\frac{\ddot{y}}{1+\dot{y}^2}-\frac{\dot{y}^2}{(1+\dot{y}^2)^2}+\frac{\dot{y}}{1+\dot{y}^2}\frac{\partial \log V}{\partial \r}-\frac{\partial\log V}{\partial y}=0,
\ee
and the Gauss-law constraint gives
\be
\tilde{F}=\frac{\tilde{Q}\o_+^{2/3}\sqrt{1+\dot{y}^2}}{\sqrt{\tilde{Q}^2 +\r^4 \o_+^{8/3}}}.
\ee
We can solve Eq.~(\ref{emod6}) if the boundary condition is given. In the $\tilde{Q} \ne 0$ case, there are fundamental strings which connect the baryonic D4 brane with the probe D6 brane. In this case, the tension of D4 and D6 branes should be balanced so the D6 embedding $\dot{y}(0)$ has a finite value so as to have an angle at the tip. This angle is determined by the force balance condition : the force at the cusp of the probe D6 brane is obtained as
\be \label{force-d6}
\hat{F}_{D6} = \frac{\partial {\cal H}_{D6}}{\partial U_c} \Bigg|_{\mbox{fix others}} = \frac{Q}{2\pi\a'}\left(\frac{1+\xi_c^{-3}}{1-\xi_c^{-3}}\right)
\frac{\dot{y}_c}{\sqrt{1+\dot{y}_c^2}}.
\ee
The whole configuration (compact D4+F1+D6) is stationary if there is a condition to balance the forces between them. The number of baryonic D4 is $Q/N_c$ since the flux from $N_C$ strings should be canceled by a single D4 brane; the ratio between the number of fundamental strings Q and $N_C$ is the number of baryonic D4. The condition which makes the system stationary ($\r=0$, $y_c =\xi_c$, the subscript c denotes the position of the cusp) is:
\be\label{force-balance2}
F_{D6}=\frac{Q}{N_c} F_{D4},
\ee
which is the ``force balance condition" (FBC), and is simplified as
\be
\dot{y}_c = \frac{\xi_c'}{y_c}.
\ee
\begin{figure} [h]
\begin{center}
    \includegraphics[angle=0, width=0.6 \textwidth]{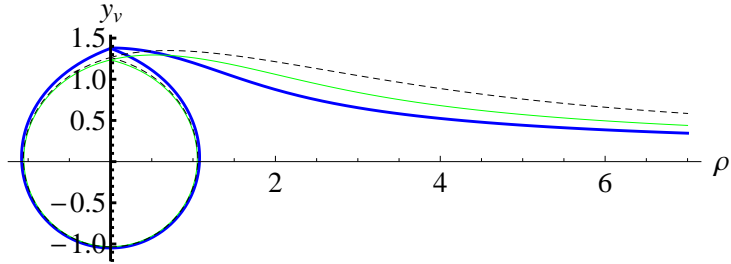}
  \caption{Left:The embedding solution of baryonic D4 and probe D6 with $y_\infty$ = 0.1 and $\tilde{Q}$ = 10$^{-4}$(thick,blue), 6.7(thin,green), 16.7(dashed,black). \label{d4d6emb1}}
\end{center}
\end{figure}
With this, the value of $\dot{y}_c$ that satisfies the FBC is uniquely determined for given values of $\xi_o$ and $m_q$. The embedding solutions of D4 and D6 are drawn in Fig. \ref{d4d6emb1} for a fixed y$_v(\infty)$ with varying $\tilde{Q}$. It is not clear why three embeddings have the same $y_\infty$, but we draw the figure only for small $\rho$ because of the lack of the space. As we can see, at the cusp, y($\rho$=0), the slope of D6 brane is different for each different $\tilde{Q}$ even for the same $y_\infty$.

\section{Pseudoscalar and Transverse Vector Meson}
As was considered in \cite{Kruczenski:2003uq}, the mesons are identified as fluctuating fields on flavor brane. Since we have a non-trivial background, the fluctuations living on the background with a non-trivial vacuum are
\ba \label{fieldconf1}
y(\rho,x^\mu) &=& y_v(\rho) + \e ~\delta y(\rho,x^\mu), \quad \phi(\rho,x^\mu) = \e ~ \delta \phi(\rho,x^\mu) \no
A_t(\rho,x^\mu) &=& \bar{A}_t(\rho) + \e~ \delta A_t(\rho,x^\mu), \quad A_x (\rho,x^\mu) = \e~ \delta A_x(\rho,x^\mu) \no
\delta y(\rho,x^\mu) &=& \exp^{-ikx}Y(\rho), \quad \delta \phi(\rho,x^\mu) = \exp^{-ikx} \Psi(\rho) \no
\delta A_t(\rho,x^\mu) &=& \exp^{-ikx}a_t(\rho) ,\quad \delta A_x(\rho,x^\mu) = \exp^{-ikx}a_x(\rho)
\ea
where $\e$ is the expansion parameter which is very small, and $\xi^2 = \rho^2 +y^2$
\be
\frac{U}{U_K} = \frac{1}{4^{1/3}}\frac{\xi}{\xi_0}\left[1+\left(\frac{\xi_0}{\xi}\right)^3\right]^{2/3}, \quad \xi_0 = \frac{U_K}{4^{1/3}}.
\ee
The embedding metric of the probe brane is
\ba
ds^2 &=& \left(\frac{U_K}{R}\right)^{3/2}U^{3/2}\eta_{\mu\nu}dx^\mu dx^\nu
+\frac{\sqrt{U_K R^3}}{\xi^2} \sqrt{U}\bigg[\Big(1+\dot{y}_v^2\Big)d\r^2 +\rho^2 d\Omega_2^2 \no
&& + \e^2 \Big(\deriv_a \delta y \deriv_b \delta y + y_v^2 \deriv_a \delta \phi \deriv_b \delta \phi \Big)dx^a dx^b + 2 ~\e~\dot{y}_v (\deriv_a \delta y) d\rho dx^a \bigg].
\ea
We plug the field configuration Eq.(\ref{fieldconf1}) into the DBI action Eq. (\ref{D6DBIaction1}) and expand it to $\epsilon^2$ order to get the linearized fluctuation. The fluctuating Lagrangian is
\ba
\calL_Y &=& \t_6 \frac{\rho^2}{\sqrt{g_0}} \bigg[\frac{ w_+^{8/3} \left(-F^2+w_+^{4/3}\right)}{2 g_0} Y'^2-\frac{2 w_+^{5/3} y_v {y_v}'\left(-4 F^2 + 3 w_+^{4/3} (1+y_v'^2)\right)}{g_0 \xi_v^5}YY' \no
&& +\frac{R^3}{\xi_0}\frac{w_+^{2/3}}{2 \xi_v^3} \left(F^2 k^2+\left(k^2-w^2\right) w_+^{4/3})\right) Y^2 \no
&& +\frac{1}{\omega_+^{2/3}g_0 \xi_v^{10}}\bigg\{ \Big(2F^2 +\omega_+^{4/3}(1+\dot{y}_v^2)\Big)\Big(F^2 -3\omega_+^{4/3}(1+\dot{y}_v^2)\Big)(1+\xi_v^3)(4y_v^2-\rho^2)\no
&& +\Big(F^2 -\omega_+^{4/3}(1+\dot{y}_v^2)\Big)\Big(2F^2 -9\omega_+^{4/3}(1+\dot{y}_v^2)\Big) -2 \omega_+^{4/3}F^2 (1+\dot{y}_v^2)y_v^2\bigg\} Y^2\bigg]
\ea
for the scalar field,
\ba
\calL_{A_t} &=&  \t_6 \frac{\rho^2}{2\sqrt{g_0}} \bigg[-\frac{w_+^{8/3}(1+y_v'^2)}{g_0} a_t'^2
+ \frac{R^3}{\xi_0} \frac{k^2 w_+^{2/3} (1+y_v'^2)}{\xi_v^3} a_t^2 \bigg]
\ea
for the longitudinal gauge field,
\be
\calL_{aY} =  \t_6 \frac{F \rho^2}{\sqrt{g_0}} \bigg[-k^2 \frac{R^3}{\xi_0} \frac{w_+^{2/3} y_v'}{ \xi_v^3} Y a_t + \frac{w_+^{8/3} y_v'}{g_0}Y' a_t' -\frac{2w_+^{1/3} y_v \left(2 F^2-w_+^{4/3} \left(1+y_v'^2\right)\right)}{g_0\xi_v^5} Y a_t' \bigg]
\ee
for the cross terms,
\ba \label{etalag1}
\calL_\phi &=&  \t_6 \frac{\rho^2 y_v^2}{2\sqrt{g_0}} \bigg[w_+^{8/3} \Psi'^2 + \frac{R^3}{\xi_0}
\frac{k^2F^2w_+^{2/3}+(w^2-k^2)w_+^2(1+y_v'^2)}{\xi_v^3}\Psi^2 \bigg]
\ea
for the Goldstone mode, and
\ba
\calL_{A_x} &=&  \t_6 \frac{\rho^2}{2\sqrt{g_0}} \bigg[w_+^{4/3} a_x'^2 + \frac{R^3}{\xi_0}
\frac{k^2 F^2 +(w^2-k^2)w_+^{4/3} (1+y_v'^2)}{w_+^{2/3} \xi_v^3} a_x^2 \bigg]
\ea
for the transverse gauge field
where
\be
g_0=w_+^{4/3}(1+y_v'^2)-F^2 = \frac{w_+^4 \rho^4(1+\dot{y}_v^2)}{Q^2+\rho^4w_+^{8/3}}.
\ee
 Note all variables are dimensionless; we just dropped tilde for all variables. The equations of motion for the Goldstone mode and the transverse vector\footnote{Here we mean by transverse vector the SO(2) transverse vector. In medium, the Lorentz symmetry, SO(1,3),  is broken, so we cannot say anything in the Lorentz covariant manner, but we still have spatial rotation SO(2). The unbroken SO(2) symmetry has transverse vectors $A_x, A_y$ when z is the direction of the wave propagation.} are
\ba \label{eomforPSVX}
0 &=& \left(\frac{\rho^2 y_v^2}{\sqrt{g_0}} \omega_+^{8/3} \Psi'\right)' -\frac{\rho^2 y_v^2}{\xi_v^3 \sqrt{g_0}}(1+y_v'^2)\omega_+^2\left[\frac{w_s^2 R^3}{\xi_0} -\frac{k_s^2 R^3}{\xi_0} \frac{\rho^4 \omega_+^{8/3}}{Q^2+\rho^4 \omega_+^{8/3}}\right]\Psi \no
0 &=& \left(\frac{\rho^2}{\sqrt{g_0}} \omega_+^{4/3} a_x'\right)'
- \frac{\rho^2}{\xi_v^3 \sqrt{g_0}}(1+y_v'^2)\omega_+^{2/3} \left[\frac{w_x^2 R^3}{\xi_0} -\frac{k_x^2 R^3}{\xi_0} \frac{\rho^4 \omega_+^{8/3}}{Q^2+\rho^4 \omega_+^{8/3}}\right]a_x .
\ea
For convenience, we will not compute the longitudinal vector and the real scalar that are coupled to each other. What we will do is to find the dimensionless eigenvalue
\be
\tilde{M}_\Psi^2 = -w_s^2 \frac{R^3}{\xi_0}=-\frac{9 w_s^2}{4^{2/3}M_{KK}^2} \quad \mbox{and}\quad \tilde{M}_{a_x}^2 = -w_x^2 \frac{R^3}{\xi_0}= -\frac{9 w_x^2}{4^{2/3}M_{KK}^2}
\ee
which is the thermal mass in dense medium when $k_s, k_x$ =0.
We impose the Neumann boundary condition at $\rho=0$ and the Dirichlett condition at $\rho=\infty$ for Eq. (\ref{eomforPSVX}): $\Psi'(0)=0  \rightarrow  \Psi(\infty)=0$ to solve the eigenvalue problem. One might wonder why we cannot impose the Dirichlett condition at $\rho=0$.
We discarded this possibility since it is unnatural to fix the $y$-position of the connection point, which can easily vibrate. The other possibility is discussed in Appendix.
And from the meson spectra for the light quark systems, i.e., $\eta^\prime$ and $\rho$ mesons  \cite{JoKimSin}, we get the Kaluza-Klein scale $M_{KK}$ = 1.039 GeV and the asymptotic distance of D4-D6 $y_\infty$. And the 't Hooft coupling constant $\lambda$ can be determined from the relation $m_q = y_v(\infty) \frac{M_{KK} \lambda}{9\pi}$.


\section{Mass vs Density}

\begin{figure} [h]
\begin{center}
    \includegraphics[angle=0, width=0.47 \textwidth]{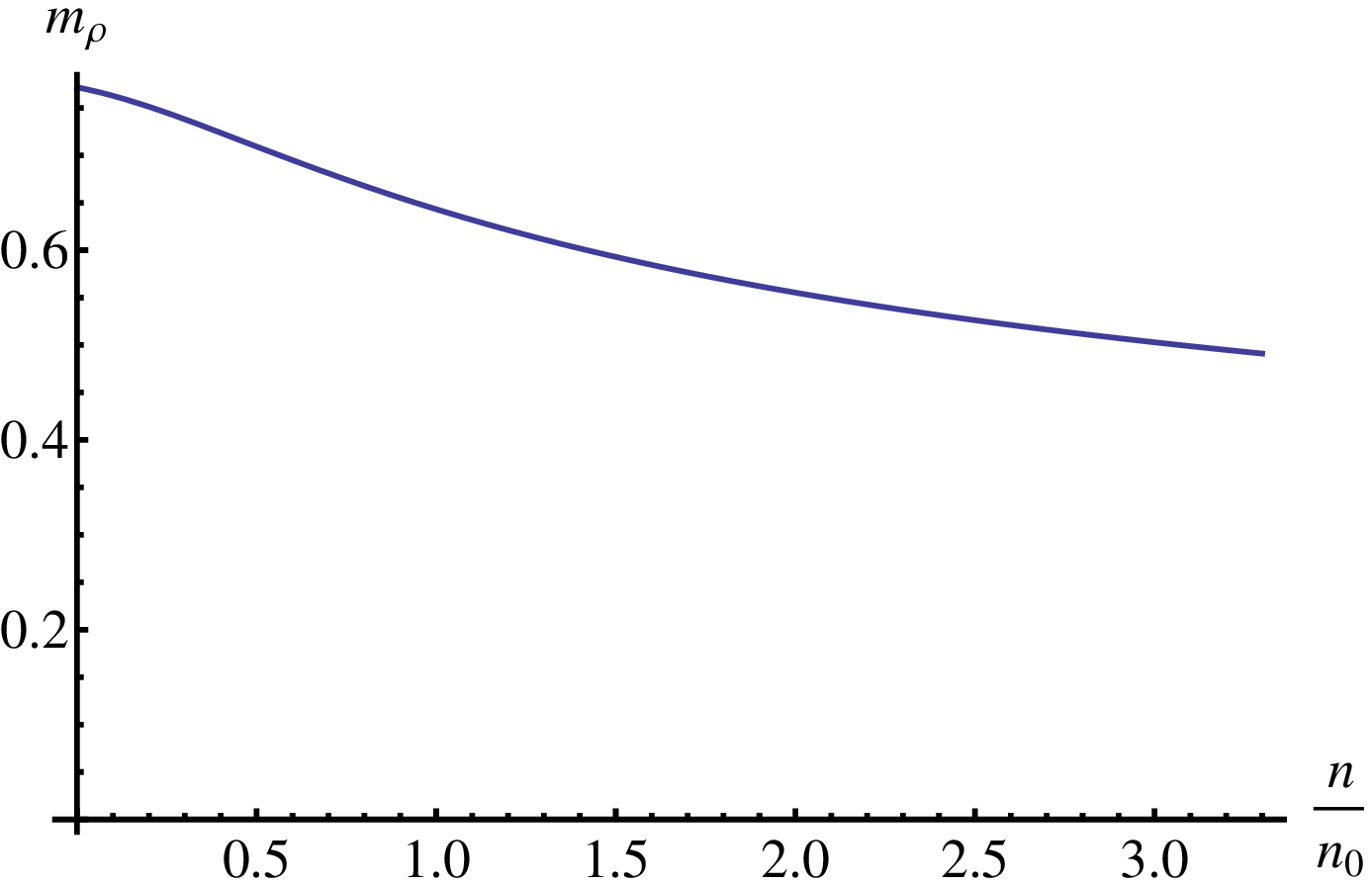}
    \includegraphics[angle=0, width=0.47 \textwidth]{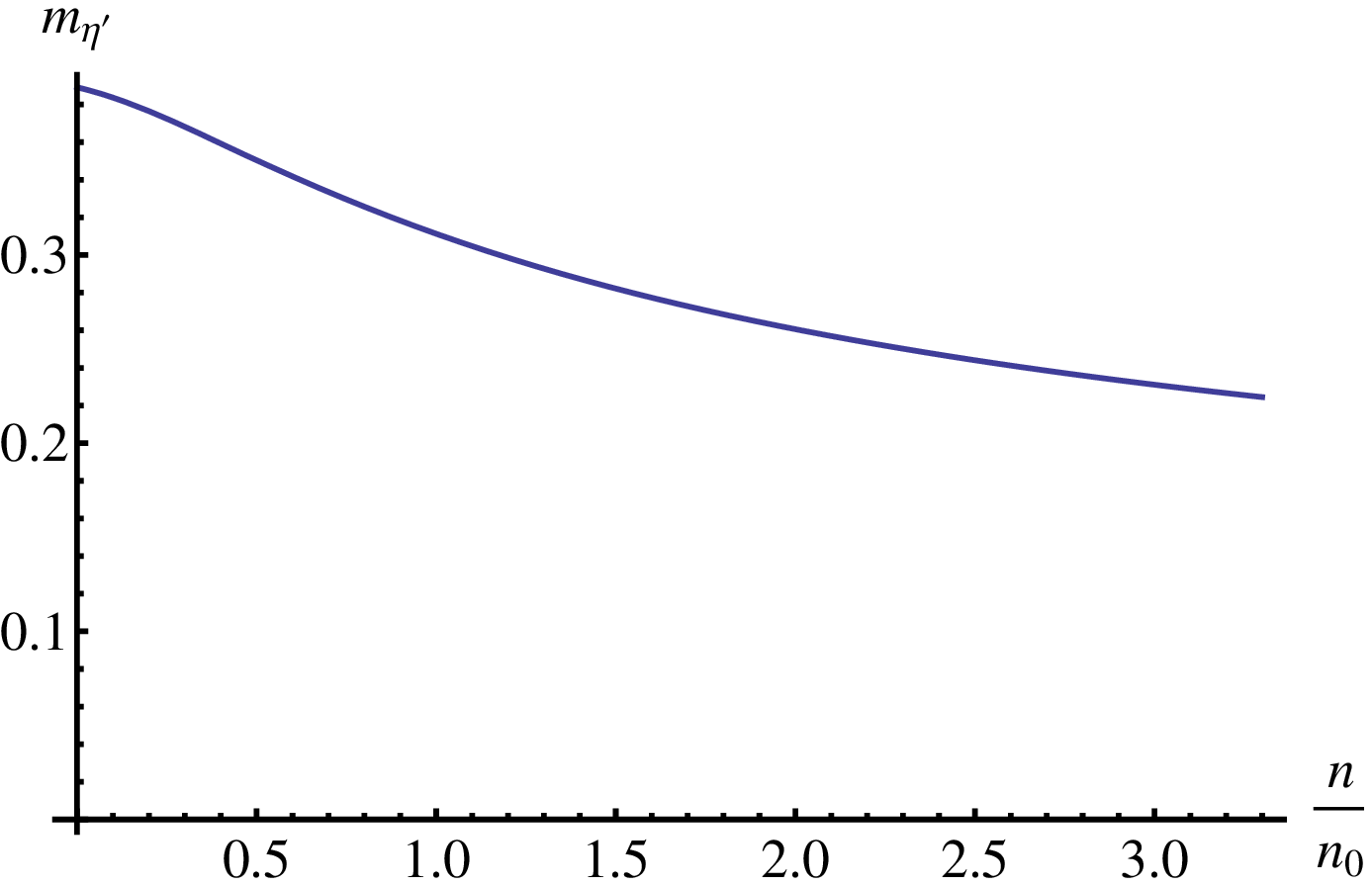}
  \caption{Left :The vector meson $\rho$ mass with $y_v(\infty)$=0.301.   Right:  The Goldstone boson $\eta^\prime$ mass.  Our result shows that the Goldstone boson mass drops slightly faster than that of the vector meson. The normal nuclear density $n_0$ was fixed by taking $\lambda = 6$.} \label{MpsivsQ1}
\end{center}
\end{figure}

The eigenvalue of Eq.~(\ref{eomforPSVX}) is obtained for varying density. For very small values of $n/n_0$,  the mass drops
slowly and then in the medium density region the mass drops almost linearly in density. Such a transition from slow to linear dropping takes place abruptly as $m_q$ decreases. For large density, the mass saturates to a finite value as seen in Fig.~\ref{MpsivsQ1}.
The linear dropping regime is a consequence of transition from concave up to concave down of the mass as a function of density.
Assuming the physical situation corresponds more or less to the linear dropping regime, we predict the mass shift of the transverse vector meson to be given by
\be
m_V(n)=m_V(0)\Phi(n)
\ee
where
\be
\Phi(n)\approx 1-0.18\cdot \frac{6}{\lambda}\frac{n}{n_0}.\label{hQCDprediction}
\ee

How does this ``prediction" compare with Brown-Rho scaling in nuclear processes? A variety of indications coming from analyses based on the scaling described in Section \ref{BRS}, i.e. Eq.(\ref{BR1}), are that the vector mesons $\rho$ and $\omega$ scale in the vicinity of $n=n_0$ as
\be
\Phi(n)\approx \frac{1}{1+y(n/n_0)} \quad \mbox{with} \quad y\approx 1/4.
\ee
This would be roughly reproduced by our model prediction Eq.~(\ref{hQCDprediction}) if $\lambda$  is  about 6. Given that our model is valid for $\lambda < N_c$, we find it to be disingenuous to fix $N_c$ = 3. In fact, there are no physical quantities in our model that depend on $N_c$ separately. Rather, they depend on the combination of $N_c$ and $g_s$. We may therefore pick the parameters within the regime where the theory is valid.
Furthermore, the saturation in the dropping mass to a finite value at high density predicted in this model (see Fig.~\ref{MpsivsQ1}) is seemingly at odds with the VM prediction Eq.~(\ref{VM}) which may be closer to nature although it has not been verified experimentally.

For the Goldstone boson, we get a similar result:
\be
m_{\eta'}(n)\approx m_{\eta'}(0)\left( 1-0.20 \cdot \frac{6}{\lambda} \frac{n}{n_0}\right),
\ee
showing that the Goldstone boson mass drops slightly faster than that of the vector meson in the linear regime. This behavior is not in agreement with what is expected in QCD. In QCD, the properties of the pion with small mass are still protected by chiral symmetry so that the mass remains more or less the same in medium as in the vacuum. This aspect will be further commented on in Section \ref{conclusion}.

One might attribute the discrepancies we have here to the nature of the probe limit of our system. At high enough density, the number of the fundamental strings are high and their gravitational effect could become significant so the background geometry could be quite different. This means that we cannot ignore the back reaction of the baryonic D4. Stated differently, what is happening is that after passing a critical density, the system becomes unstable but there will be a stable geometry with full back-reaction of the baryonic D4 and the flavor D6. This should correspond to ``beyond quenched approximation" in lattice gauge theory. As is shown in \cite{sin}, the gravity back reaction of charge can restore the possibility of phase transition driven by high density.

One can now see the geometric reason why the mass has to decrease; since the effective tention $\tau_6^{eff}(\rho)= \tau_6 V(\rho)$ in Eq.~(\ref{effectiveTension}) increases with increasing $\tilde{Q}$, it costs energy to bend the brane. Note that the fluctuation wave function describes the deformation of the brane. With our boundary condition $\Psi'(0)=0$ the wave function becomes smoother for large $\tilde{Q}$ and the energy for this less-bending configuration should be lowered. Therefore the large $\tilde{Q}$, the smaller is the meson mass. This feature is captured in Fig.~\ref{phiWavefunction} in Appendix B.

\section{The ``Pion" ($\eta^\prime$) Velocity}
\begin{figure} [h]
\begin{center}
    \includegraphics[angle=0, width=0.45 \textwidth]{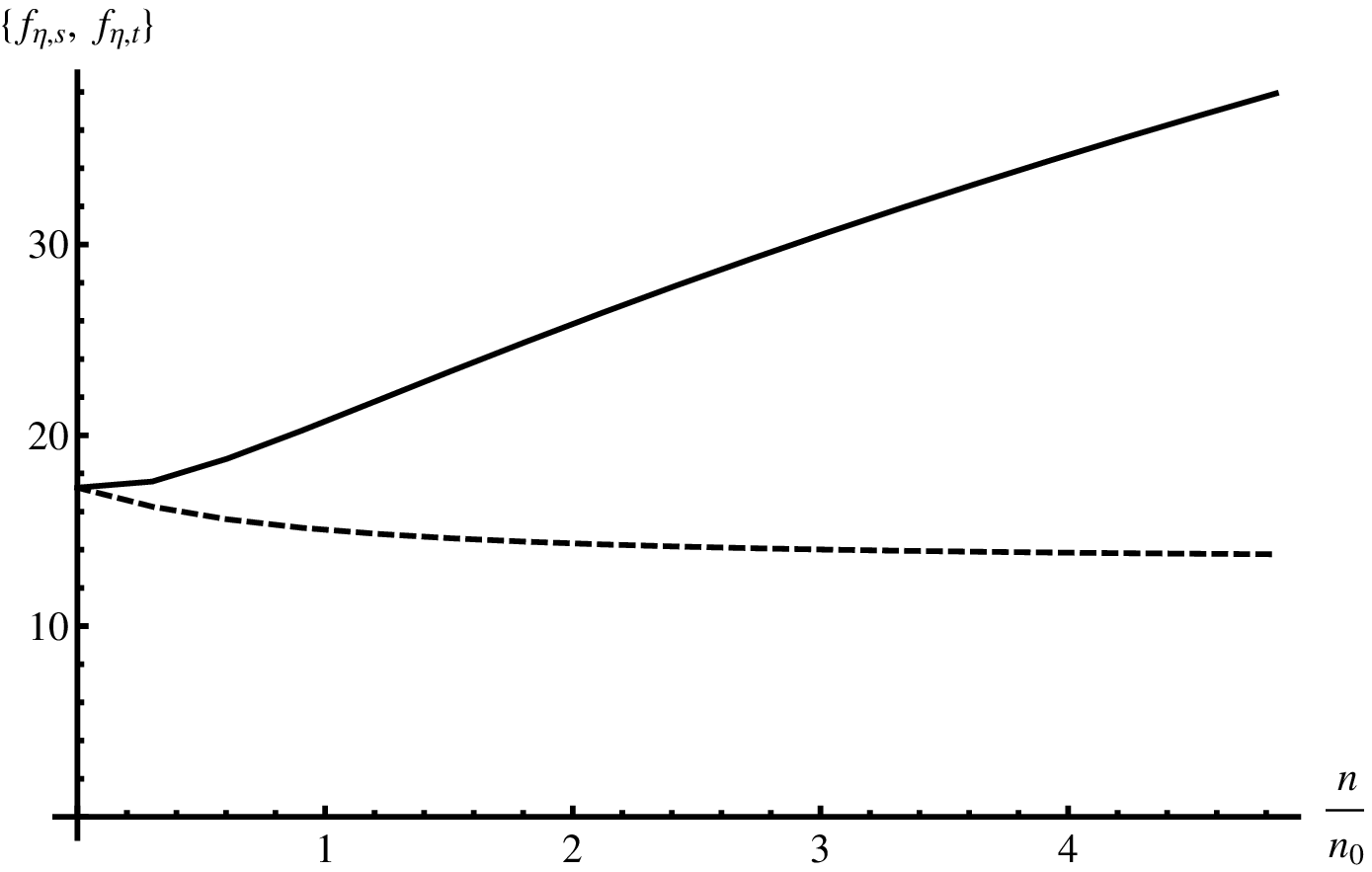}
    \includegraphics[angle=0, width=0.45 \textwidth]{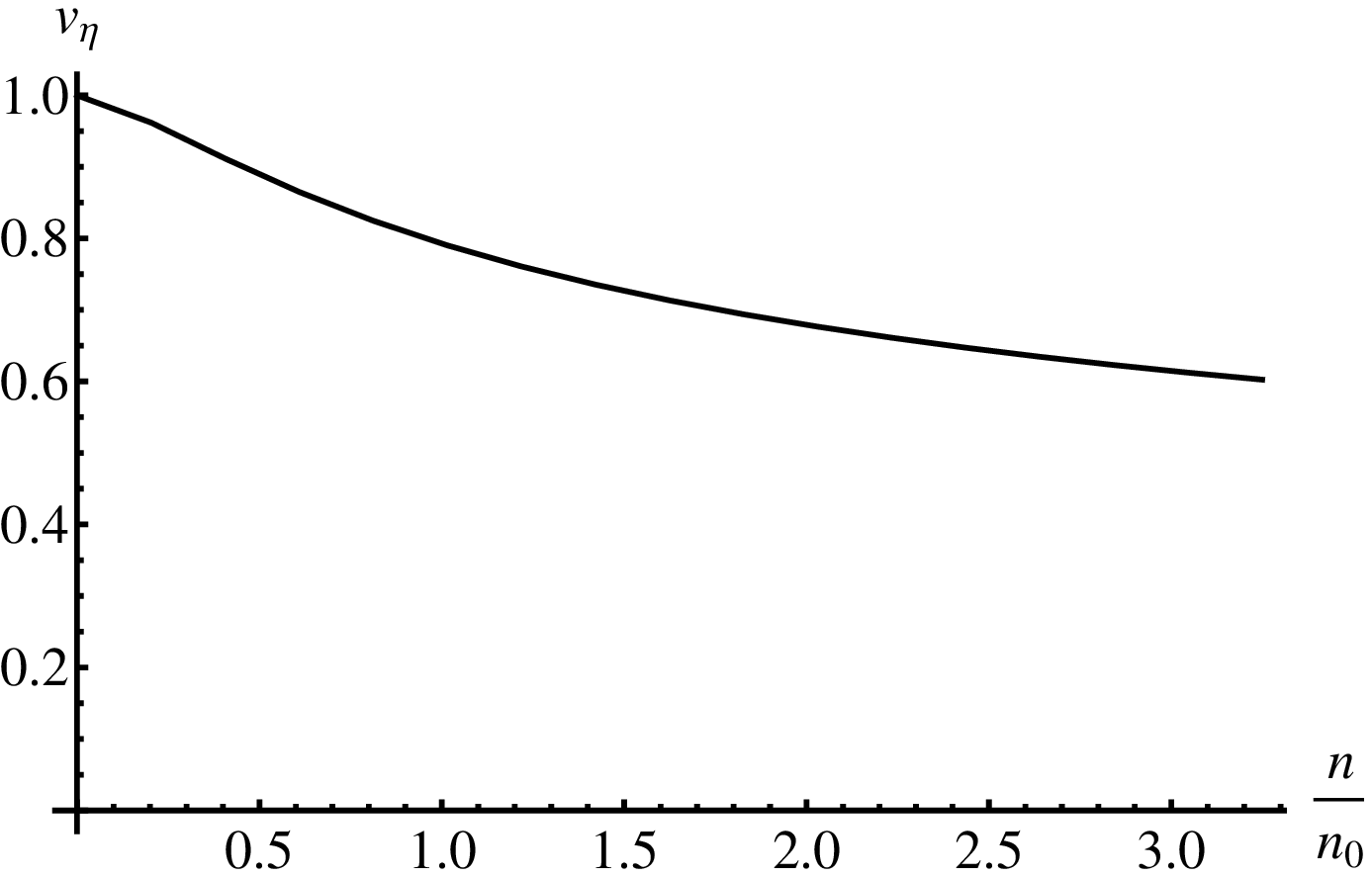}
  \caption{The pion decay constants are given as a function of density. The temporal (spatial) decay constant increases (decreases) from 17 MeV to 21(15) MeV at the normal nuclear density. The pion velocity is presented as a function of density in the right panel. The pion velocity decreases when density is increased; at the normal nuclear density, $n=n_0$, the pion velocity $v_\pi = \sqrt{\frac{f_\pi^s}{f_\pi^t}}$ is about 0.85. \label{pdeccons} }
\end{center}
\end{figure}

The pion velocity\footnote{Here we use ``pion" as the pseudo-Goldstone mode, not the real pion meson. In our system, pion corresponds to the $\eta^\prime$ meson. In QCD, the $\eta^\prime$ is pushed up in mass by the $SU(1)_A$ anomaly which is the $1/N_c$ effect. In the large $N_c$ limit, it is degenerate with the pion.} is defined as the ratio between the temporal and the spatial pion decay constants, $f_\pi^s/f_\pi^t$. From Eq.~(\ref{etalag1}), we define the pion decay constant as a wave-function renormalization factor of $\phi$,
\ba
\calL_{4d, \phi} &=& \frac{f_{\pi,t}^2}{2} \int dt d^3 x  \deriv_t \phi \deriv^t \phi+\frac{f_{\pi,s}^2}{2} \int dt d^3 x  \deriv_i \phi \deriv^i \phi \no
(f_\pi^t)^2 &=& \frac{\tau_6 R^3}{V_3 \xi_0} \int_0^\infty d\rho \frac{\rho^2 y_v^2}{\xi_v^3\sqrt{g_0}}\omega_+^2 (1+\dot{y}_v^2) \Psi(\rho)^2, \quad
(f_\pi^s)^2 = \frac{\tau_6 R^3}{V_3 \xi_0} \int_0^\infty d\rho \frac{\rho^2 y_v^2}{\xi_v^3}\omega_+^{2/3}\sqrt{g_0}\Psi(\rho)^2.
\ea
The numerical results obtained from these formulae are given in Fig.~\ref{pdeccons}. One sees that
the spatial pion decay constant decreases while the temporal decay constant increases as density increases. As a result, the pion velocity decreases to almost 0.6 at the normal nuclear density. This is similar to the result obtained in \cite{pionvelocity}
from the Sakai-Sugimoto model\cite{SS}. Our present result shows more rapid drop of the pion velocity since the spatial part of the
pion decay constant does not increase at all. In QCD-motivated effective theories, one expects that the temporal pion decay constant decreases, not increases, in density and hence there is again a difference from the gauge theory sector. This aspect will also discussed in Section \ref{conclusion}.

\section{Density-Dependent Dispersion Relation}
\begin{figure} [h]
\begin{center}
    \includegraphics[angle=0, width=0.31 \textwidth]{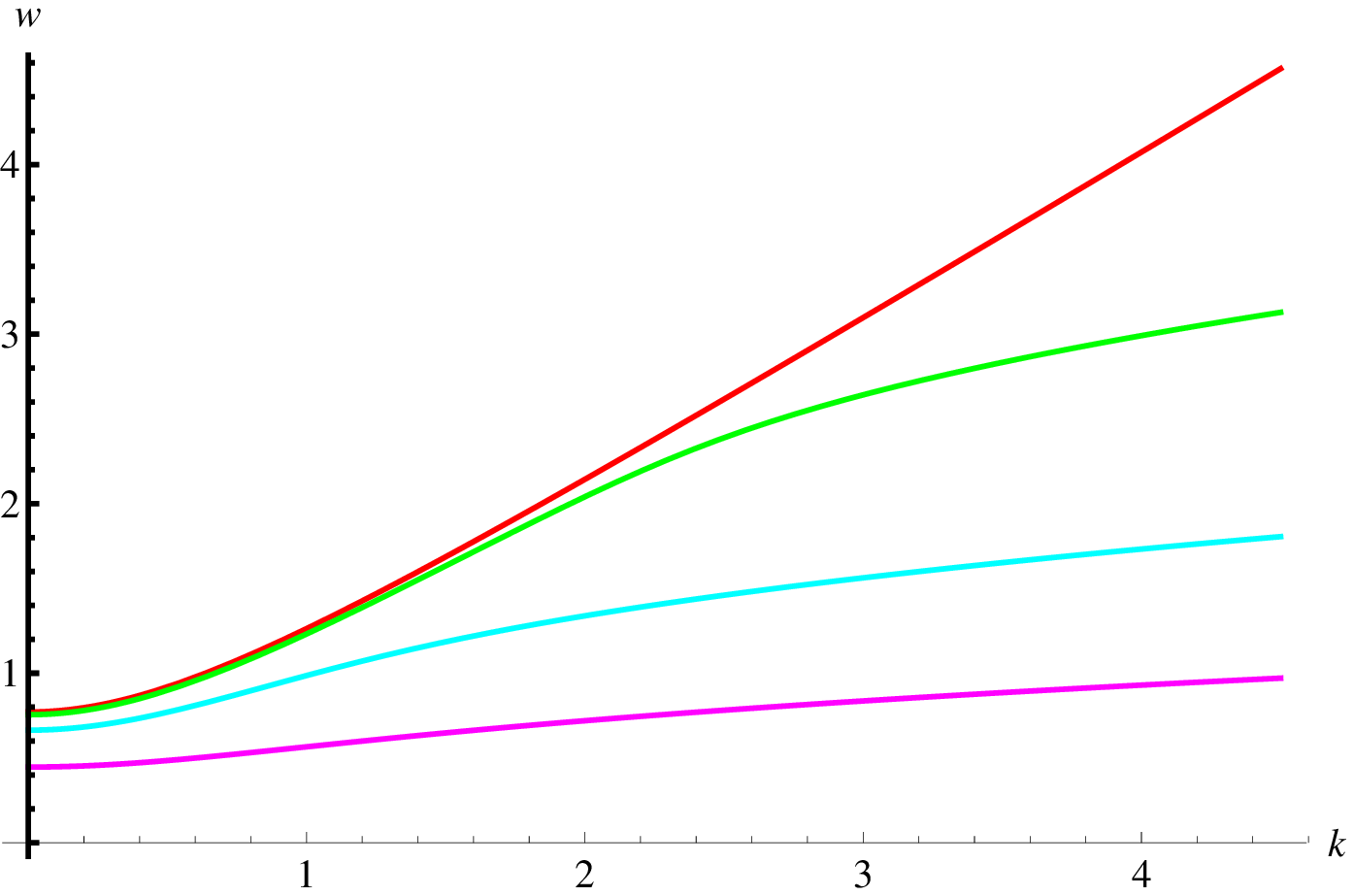}
    \includegraphics[angle=0, width=0.31 \textwidth]{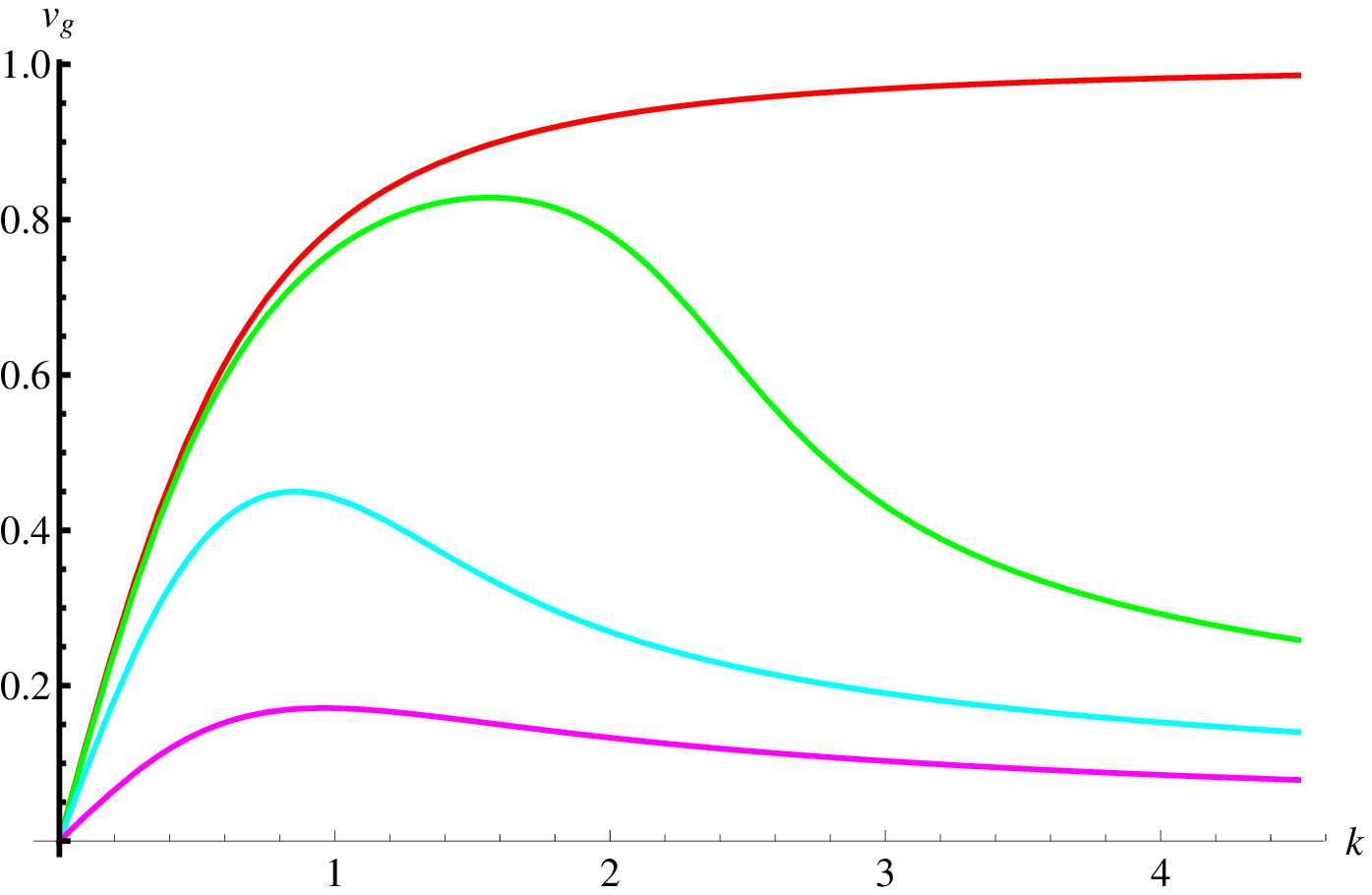}
    \includegraphics[angle=0, width=0.31 \textwidth]{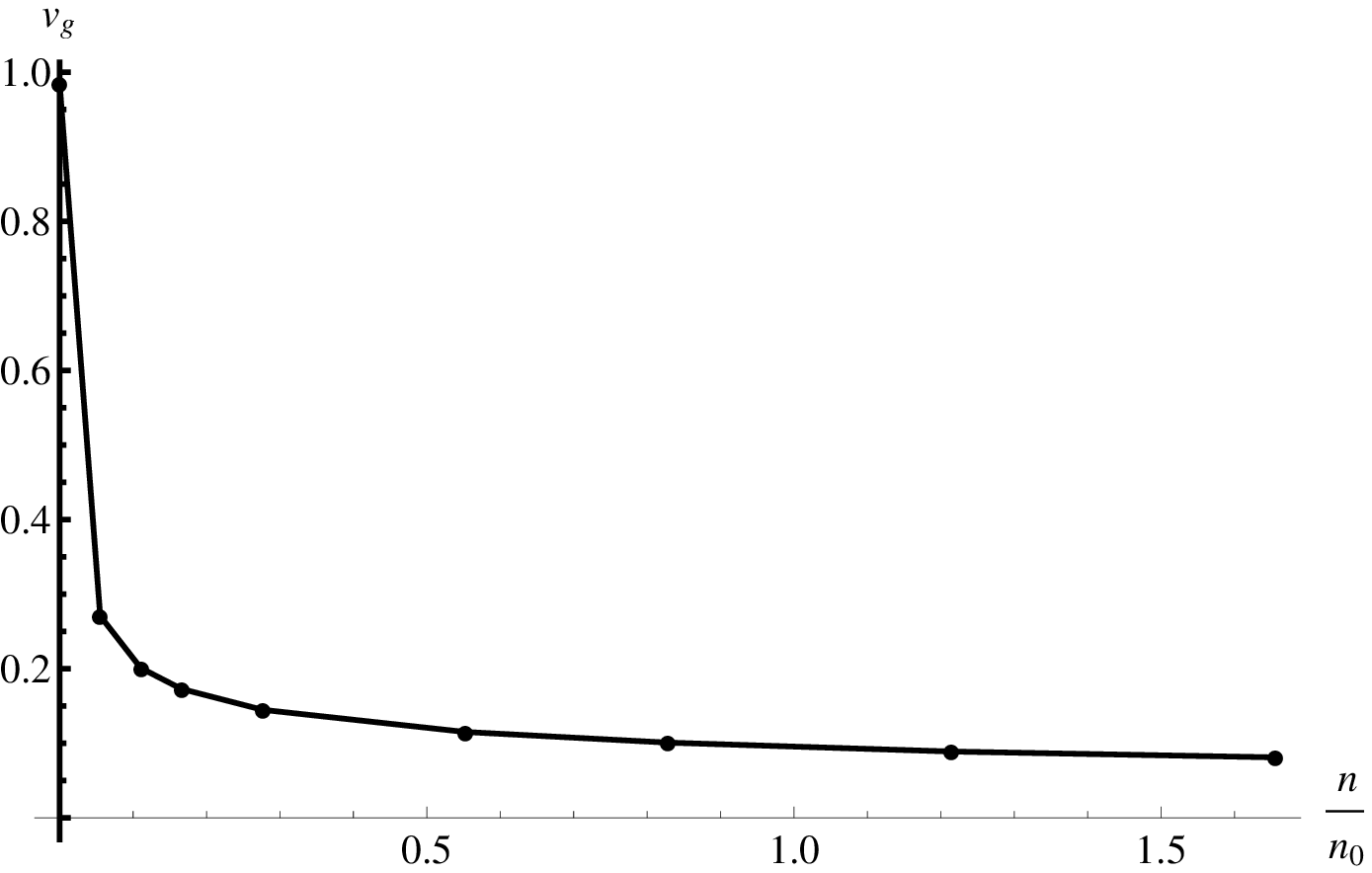}
  \caption{Left panel: The dispersion relation for the transverse vector with $y_v(\infty)$ =0.303. The eigenvalue $w$ is plotted. The solid lines correspond to $\tilde{Q}$ = 0 (red), .2(green), 1(skyblue), 6(pink) as a function of $k$. Middle panel: $\frac{dw}{dk}$ as a function of $k$. Right panel: The group velocity $v_g = \frac{d w(k)}{d k}|_{k\rightarrow \infty}$ is given and for each line. \label{dispersionrelationofetap} }
\end{center}
\end{figure}

The dispersion relation is the relation between the frequency $w$ and the spatial momentum $k$. And the frequency is numerically computed as the eigenvalue of Eq.~(\ref{eomforPSVX}) by varying the parameter $k$\footnote{We compute $-w^2$ in Eq.~(\ref{eomforPSVX}) by varying the parameter $k^2$ because of the sign convention $-M^2 =-w^2+k^2$.}. Fig. \ref{dispersionrelationofetap} shows the dispersion relation of the pseudo-scalar (left) and the group velocity (right). At zero density  the dispersion relation
is  $w^2 - k^2 = M^2$. What we want to see is how this relation is modified in dense medium.
Given the dispersion relation, the group velocity is calculated by $\frac{d\omega}{dk}$.  In our approach,  the dispersion relation has two interesting features. First there is a sudden decrease of group velocity in the intermediate $k$ region. The group velocity is linear in $k$ when $k$ is less than the rest mass(0.77 GeV for $\rho$ meson) and $v_g$ is a constant when $k$ is large compared to the mass. But when $k$ is larger than 1, the group velocity suddenly decreases and saturates to a certain value, Fig \ref{dispersionrelationofetap}. Secondly, the group velocity is not 1 in our dense medium (compare the zero density case, red line in Fig.~\ref{dispersionrelationofetap}). It slowly decreases when baryon density is increased.

\section{Remarks and Conclusion}\label{conclusion}
We have solved the embedding function of baryonic D4 and probe D6 with FBC (force balance condition) with the embedding parameterized by $y_v(\infty)$ and $\tilde{Q}$. The density dependence of the meson mass obtained therefrom is studied for the same $y_v(\infty)$. We find no qualitative difference between the Goldstone mode and the transverse vector mode: their mass drops substantially, at the normal nuclear density but does not vanish at high density. Thus chiral symmetry does not get restored at any density. Both are at odds with what one expects in QCD: The pion mass, protected by chiral symmetry (though lightly broken), should not change much by density and chiral symmetry should get restored at least in the chiral limit at large density.

As mentioned in Introduction, whether the vector-meson mass can be taken as an order parameter of chiral symmetry to be probed in physical observables depends on whether local fields are relevant degrees of freedom at the phase change driven by density. In the vicinity of nuclear matter density where the quasiparticle picture should be applicable, hadron masses, both nucleon and meson, do seem to drop proportionally to the pion decay constant (more precisely the time component $f_\pi^t$). The drop is roughly 20\%. Our model agrees with the BRS/VM scenario when we take 't Hooft coupling $\lambda$ =6. The value of $\lambda$ needed to fit the mass drop near $n_0$ is around 6, which exceeds 2.5, the bound for the validity of the scheme for $N_c$=3. The number of colors $N_c$, however, figures nowhere in our analysis and hence need not be fixed to any value. Hence taking $\lambda$=6 does not spoil the validity of our model.  

It should be stressed, however, that it is not obvious that the system could be described in terms of quasiparticles in the vicinity of the phase change. The system could be totally ``melted."  Up to day, searches for the evidence of vanishing vector meson mass in experiments have not yet come out positive. Therefore our hQCD result that the vector meson mass cannot vanish is not ruled out by nature. It remains to see what happens when $1/N_c$ and $1/\lambda$ corrections are made to the model.

Let us consider the pion velocity for which there are results in effective field theories in the QCD sector. Effective field theory encompassing the current algebra relations and spectral decomposition and using linear density approximation predicts that the Gell-Mann-Oakes-Renner relation -- which is satisfied in the matter-free vacuum by the D4-D6 model used here -- is given in medium in terms of the time component of the pion decay constant $f_\pi^t$ as~\cite{hatsuda-kunihiro}
\bea \label{GMOR1}
(f_\pi^t/f_\pi)^2 (m_\pi^*/m_\pi)^2=\la\bar{q}q\ra^*/\la\bar{q}q\ra
\eea
which implies that as $\la\bar{q}q\ra^*$ drops at increasing density, the $f_\pi^t$ will also decrease. The same line of arguments indicate that the pion mass is largely protected by chiral symmetry so that it changes little in medium. What we found in this hQCD model seems not to be following the QCD-anchored intuition.

\begin{figure} [h]
\begin{center}
    \includegraphics[angle=0, width=0.5 \textwidth]{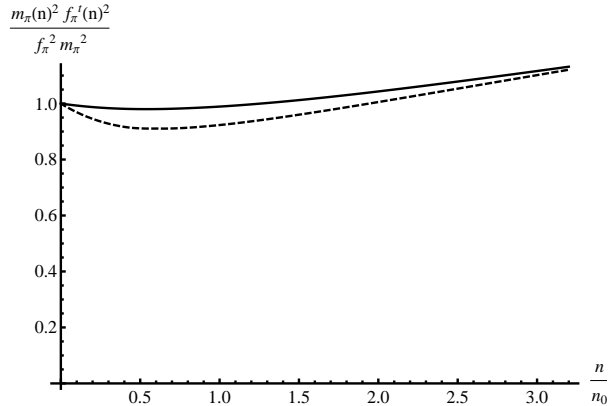}
  \caption{The graphical presentation of the GMOR relation is given as a function of density. The dashed line is for $(f_\pi^t/f_\pi)^2 (m_\pi^*/m_\pi)^2$ and the solid line for the fermion condensate, $\frac{\la\bar{\psi}\psi\ra(n)}{\la\bar{\psi}\psi\ra}$. At $n_0$ with 't Hooft constant $\lambda$=6, $\frac{\la\bar{\psi}\psi\ra(n_0)}{\la\bar{\psi}\psi\ra}$=0.99.. \label{ccvsn}}
\end{center}
\end{figure}

It is intriguing and perhaps significant, however, that in this hQCD model, the left hand side (LHS) and the right hand side (RHS) of Eq.~(\ref{GMOR1}) are balanced even though $(f_\pi^t/f_\pi)^2$ and $\la\bar{q}q\ra^*/\la\bar{q}q\ra$ separately behave differently, as one can see in Figs.~\ref{pdeccons} and \ref{ccvsn}. In our model, $\la\bar{q}q\ra^*/\la\bar{q}q\ra\approx 0.99$, $f_\pi^t/f_\pi\approx 21/17$ and $m_\pi^*/m_\pi\approx 0.8$ at n=n$_0$. So in Eq.~(\ref{GMOR1}) LHS $\approx 0.95$ and RHS $\approx  0.99$. The GMOR relation appears to hold even at finite density -- as it does in matter-free space -- in our hQCD model. So if there were any defect in one, the same would be present in the other.

What can be happening with the BRS/VM can be understood more clearly in hot matter. Suppose that the relevant low-mass degrees of freedom near chiral restoration are the pions only. In this case, one can use chiral effective Lagrangian field theory to compute near the critical temperature the pion decay constant $f_\pi^t$ -- which is found to be equal to the axial susceptibility $\mu_A$ -- which is not zero, whereas the space component $f_\pi^s$ tends to zero, hence finding $v_\pi\rightarrow 0$~\cite{Son}. This result is drastically modified when the vector meson mass drops to zero (in the chiral limit) as in BRS/VM~\cite{hkr}. In this case, both ${f_\pi^s} $ and $f_\pi^t$ drop to zero, with the $v_\pi$ going to 1. What happens is that the the zero-mass vector meson cancels the pion contribution in $f^t_\pi$. The situation is not as clear in density as in temperature. However simulating dense matter with skyrmions -- which is essentially the same as the BRSA --  shows that $v_\pi$ behaves similarly in density~\cite{BYP}.


The glaring differences between the the two -- hQCD and QCD-motivated -- approaches raise the big question as to where the defects are and how to remedy them. While the QCD-motivated approaches are found to be consistent with observations up to the density probed by experiments, going beyond that density remains unverified and hence could very well be incorrect due to the lack of reliable theoretical tools to handle strong-coupling regimes. On the other hand, the hQCD-based models -- seemingly inconsistent with available observations in hadron physics -- can more readily access the strong-coupling regime within certain approximations.  Connecting the two approaches will likely provide the means to unravel the structure of the poorly understood cold dense matter system. To improve the situation for hQCD we first have to consider the back-reaction of the gravity to the probe brane, that is, the gluon-fermion interaction  effect. The situation is quite similar in the Sakai-Sugimoto model.
It is possible that this increasing tendency in both the pion decay constant and the quark condensate for increasing density is due to the
large $N_c$ nature. In that case, we need to consider the finite $N_c$ effect, which is beyond the gravity approximation.
The result on the pion velocity $v_\pi$ presents also a conflict between the dual descriptions. The result found in the previous section is that at increasing density, the pion velocity decreases from 1 in the vacuum to near zero at high density. Model calculations based on the BRS/VM modeling QCD indicate the contrary, that is, the $v_\pi$ tends to go back to 1 near the chiral transition point.

\subsection*{Acknowledgments}
This work was supported by the WCU project of Korean Ministry of Education, Science and Technology (R33-2008-000-10087- 0). This work is also supported by the Mid-career Researcher Program through NRF Grant No. 2010-0008456 and by the National Research Foundation of Korea (NRF) grant funded by the Korea government (MEST) (No. 2005-0049409).

\vskip 1cm
\appendix
\centerline{\bf\Large Appendix}
\setcounter{equation}{0}
\renewcommand{\theequation}{A\arabic{equation}}
\section{WZ term}
Taking from Chapter 19 of \cite{Ortin} the type-IIA super gravity solution of D4 brane
\ba
ds^2 &=& \frac{1}{\sqrt{H_{D4}}}\left[-Wdt^2 +d\vec{x}^2\right]+\sqrt{H_{D4}}\left[\frac{d\rho^2}{W}+\rho^2d\Omega_4^2\right] \no
\exp^{-2\hat{\phi}} &=& \exp^{-2\hat{\phi}_0} \sqrt{H_{D4}}, \quad
\hat{C}^{(5)}_{ty^1y^2y^3y^4}=\alpha \exp^{-\hat{\phi}_0} (H_{D4}^{-1}-1)\no
H_{D4}&=&1+\frac{h_{D4}}{\rho^3}, \quad W = 1+\frac{h_{D4} (1-\alpha^2)}{\rho^3} ,\quad h_{D4}=\frac{g(2\pi l_s)^4}{3\omega_4},
\ea
the near horizon limit $H_{D4}=\frac{h_{D4}}{\rho^3}$, and renaming the coordinates, we write the non-extremal solution as
\ba
ds^2 &=& \sqrt{\frac{U^3}{R^3}}\left[-fdt^2 +d\vec{x}^2\right]+\sqrt{\frac{R^3}{U^3}}\left[\frac{dU^2}{f}+U^2d\Omega_4^2\right], \quad \exp^{-\phi} = \frac{1}{g_s}\left(\frac{R^3}{U^3}\right)^{1/4} \no
f &=& 1-\frac{U_K^3}{U^3}, \quad \hat{C}^{(5)}_{tx^1x^2x^3x^4}=-\frac{1}{g_s} \sqrt{1+\frac{U_K^3}{R^3}} \left(1-\frac{U^3}{R^3}\right)
\ea
where the parameters are identified as
\be
\rho = U,\quad h_{D4}=R^3, \quad \exp^{\hat{\phi}_0}=g_s,\quad \alpha^2 = 1+\frac{U_K^3}{h_{D4}}.
\ee
The five form field has its six form field strength, $dC^{(5)}$
\be
dC^{(5)}_{tx^1x^2x^3x^4U} = \frac{1}{g_s} \sqrt{1+\frac{U_K^3}{R^3}} 3\frac{U^2}{R^3}
\ee
and its magnetic dual is
\ba
C^{(4)}&=&\star\left(dC^{(5)}\right)_{\theta\psi_1\psi_2\psi_3} = \sqrt{g}g^{tt}g^{11}g^{22}g^{33}g^{44}g^{UU}dC^{(5)}_{tx^1x^2x^3x^4U}\no
&=&\frac{1}{g_s} \frac{R^6}{U^2}\sin^3\theta ~\Omega_3~ \sqrt{1+\frac{U_K^3}{R^3}} 3\frac{U^2}{R^3} = 3\frac{R^3}{g_s} \sin^3\theta~\Omega_3~\sqrt{1+\frac{U_K^3}{R^3}}.
\ea
Note that the volume of $S^n$ is
\ba
d_{S^{n-1}}V &=& \mbox{sin}^{n-2}(\phi_1) \mbox{sin}^{n-3}(\phi_2)\cdots \mbox{sin}(\phi_{n-2}) d\phi_1 \cdots d\phi_{n-1} \no
d\Omega_4 &=& \sin^3 \theta d\Omega_3.
\ea

\section{Other boundary condition}
\setcounter{equation}{0}
\renewcommand{\theequation}{B\arabic{equation}}
In our treatment given in the main text, we used the Neumann boundary condition (hereafter N). Here let us consider the Dirichlet boundary condition (hereafter D) in computing the meson spectrum
\be
\Psi(0)=0 \qquad \longrightarrow \qquad \Psi(\infty)=0.
\ee

\begin{figure} [h]
\begin{center}
    \includegraphics[angle=0, width=0.35 \textwidth]{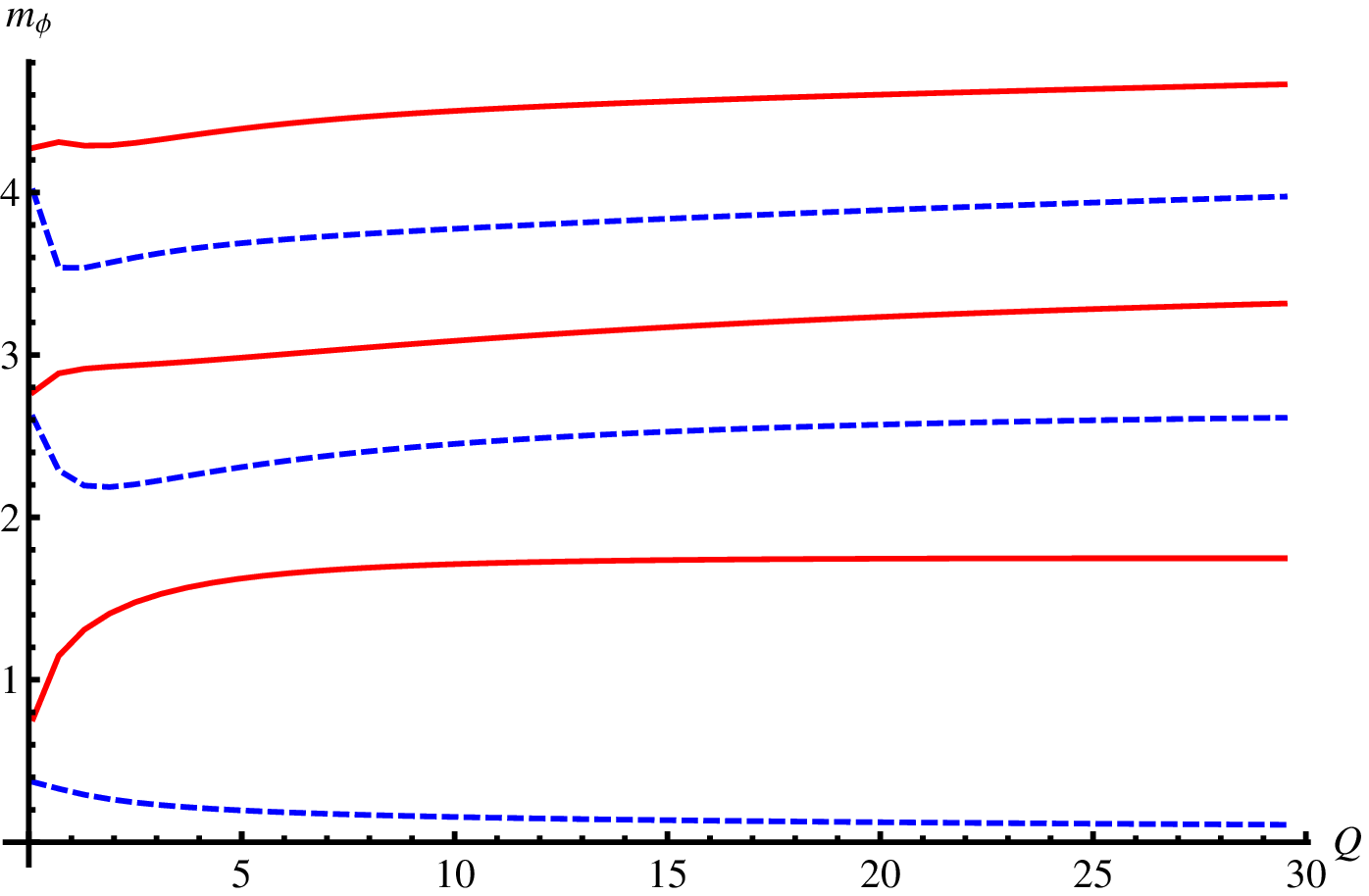}
    \includegraphics[angle=0, width=0.35 \textwidth]{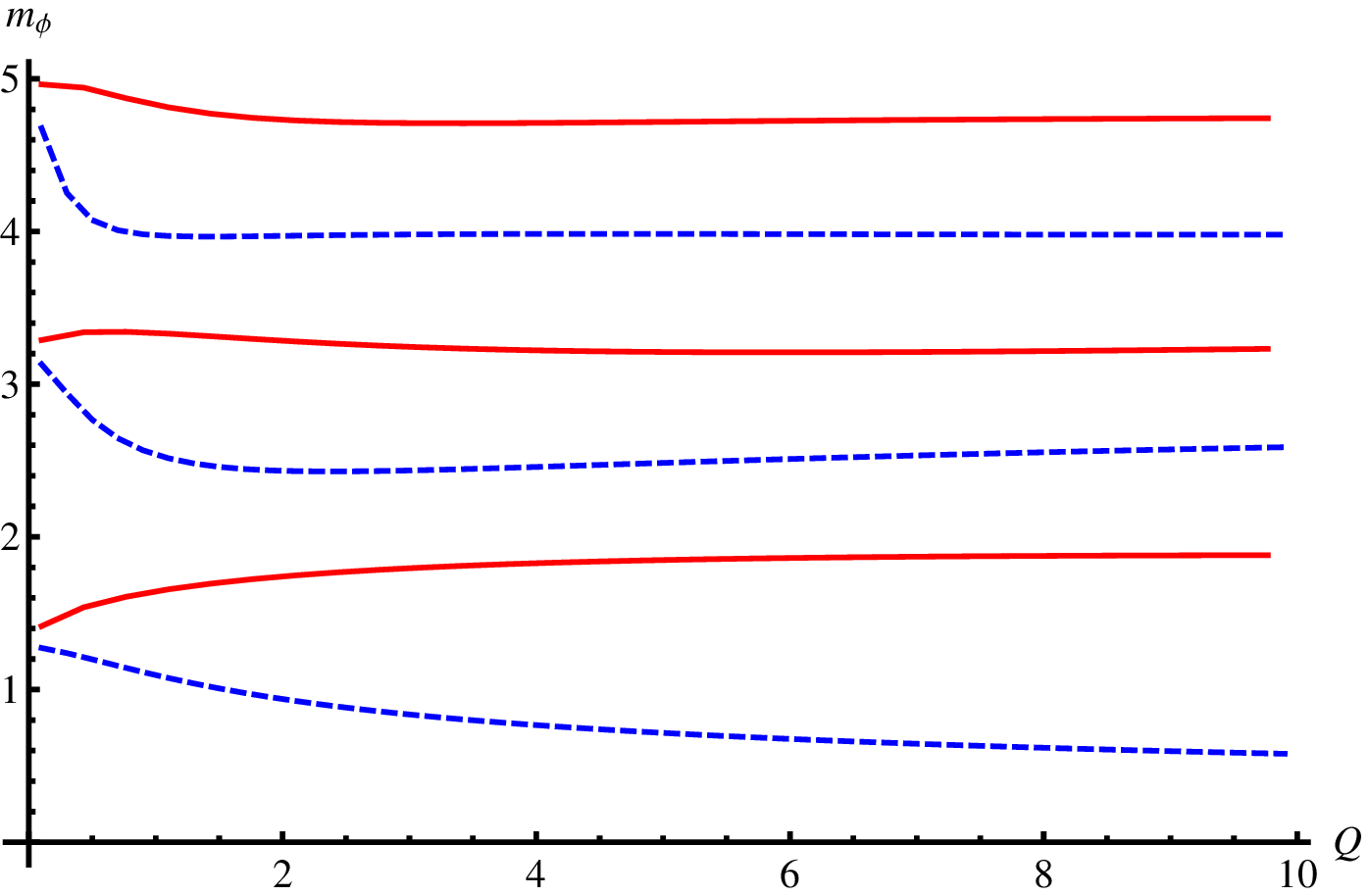}
    \includegraphics[angle=0, width=0.35 \textwidth]{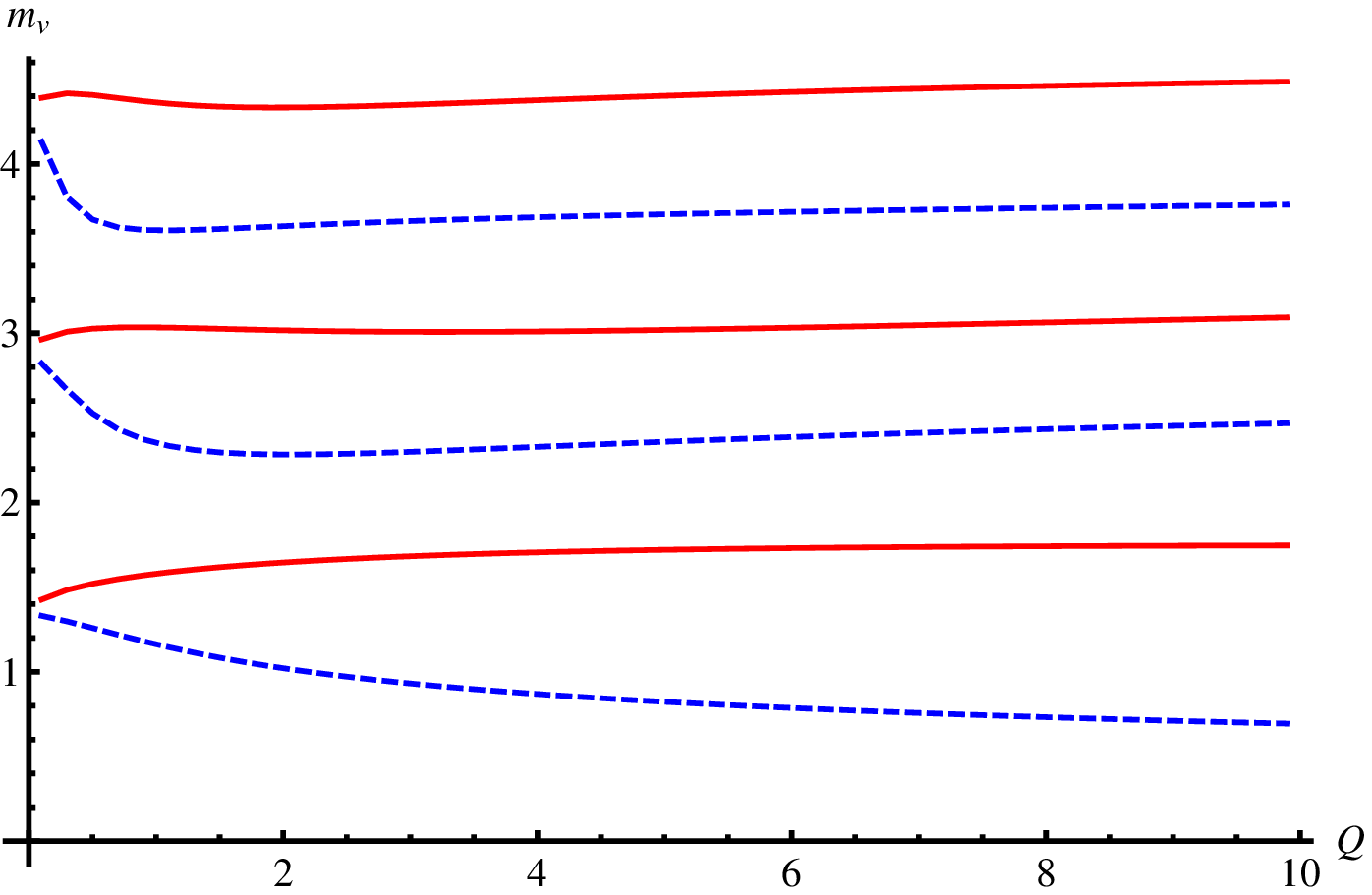}
    \includegraphics[angle=0, width=0.35 \textwidth]{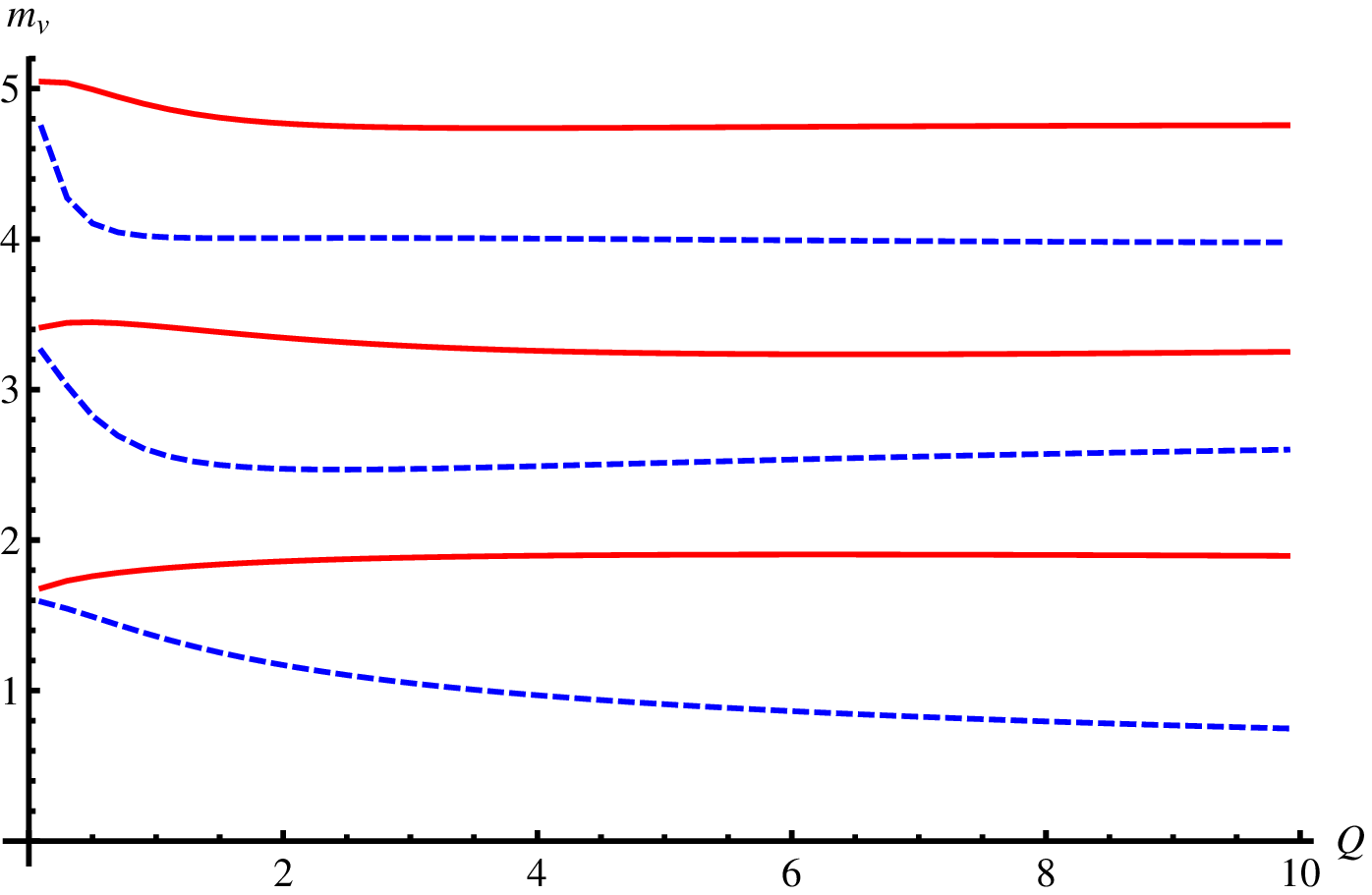}
  \caption{The meson masses is presented as a function of the dimensionless density, $\tilde{Q}$. The red thick line is of the Dirichelet boundary condition and the blue dashed line is of the Neumann.
  TOP - The masses of lowest and first two excitations of $\eta \prime$ are plotted from bottom. Top-left figure is of $y_\infty$=0.1 and top-right is of $y_\infty$ = 1. BOTTOM - The masses of lowest and first two excitations of transverse vector meson are plotted from bottom and the same parameters are used as top figure. \label{mphimqwithDN1}}
\end{center}
\end{figure}

\begin{figure} [h]
\begin{center}
    \includegraphics[angle=0, width=0.4 \textwidth]{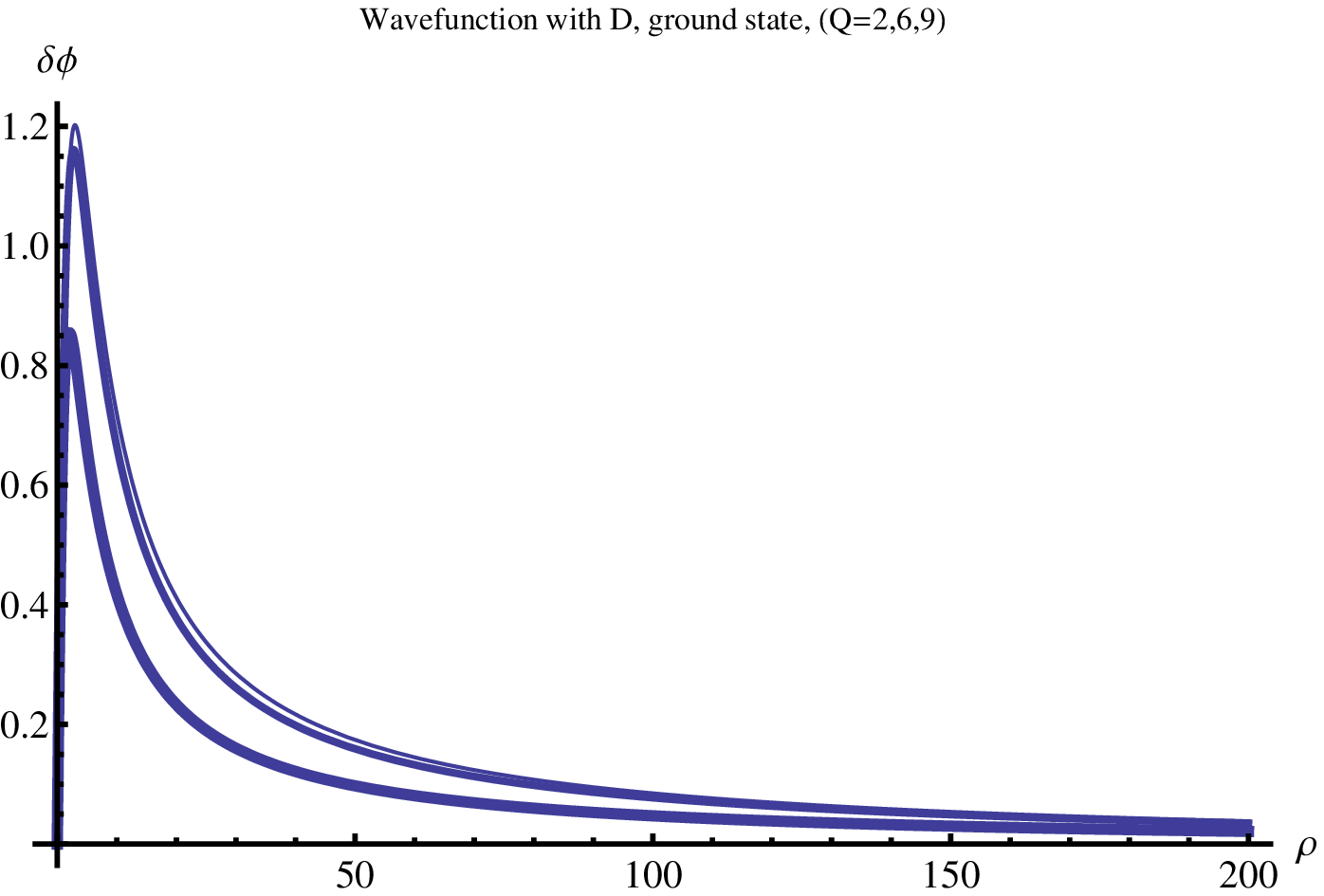}
    \includegraphics[angle=0, width=0.4 \textwidth]{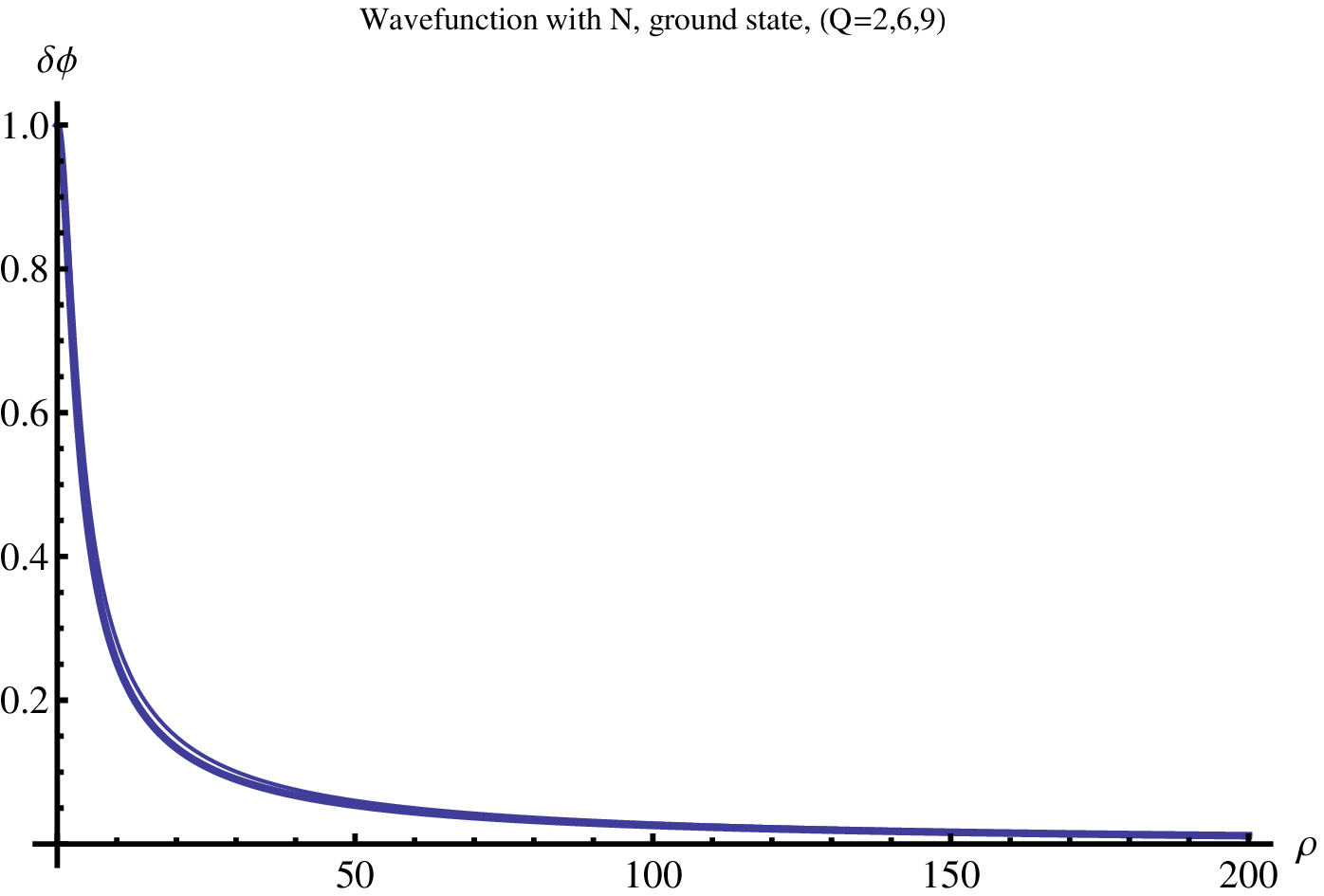}
    \includegraphics[angle=0, width=0.4 \textwidth]{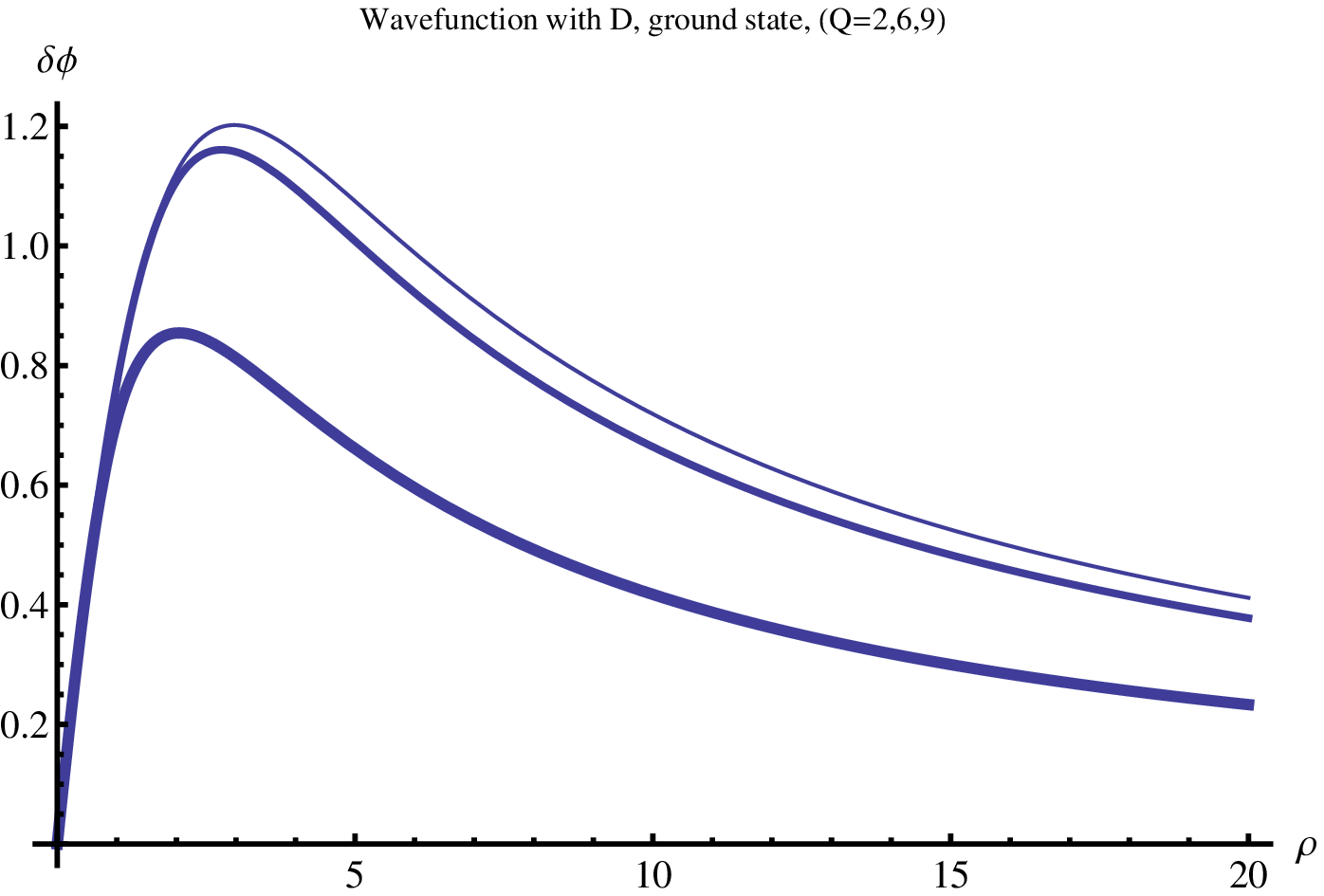}\
    \includegraphics[angle=0, width=0.4 \textwidth]{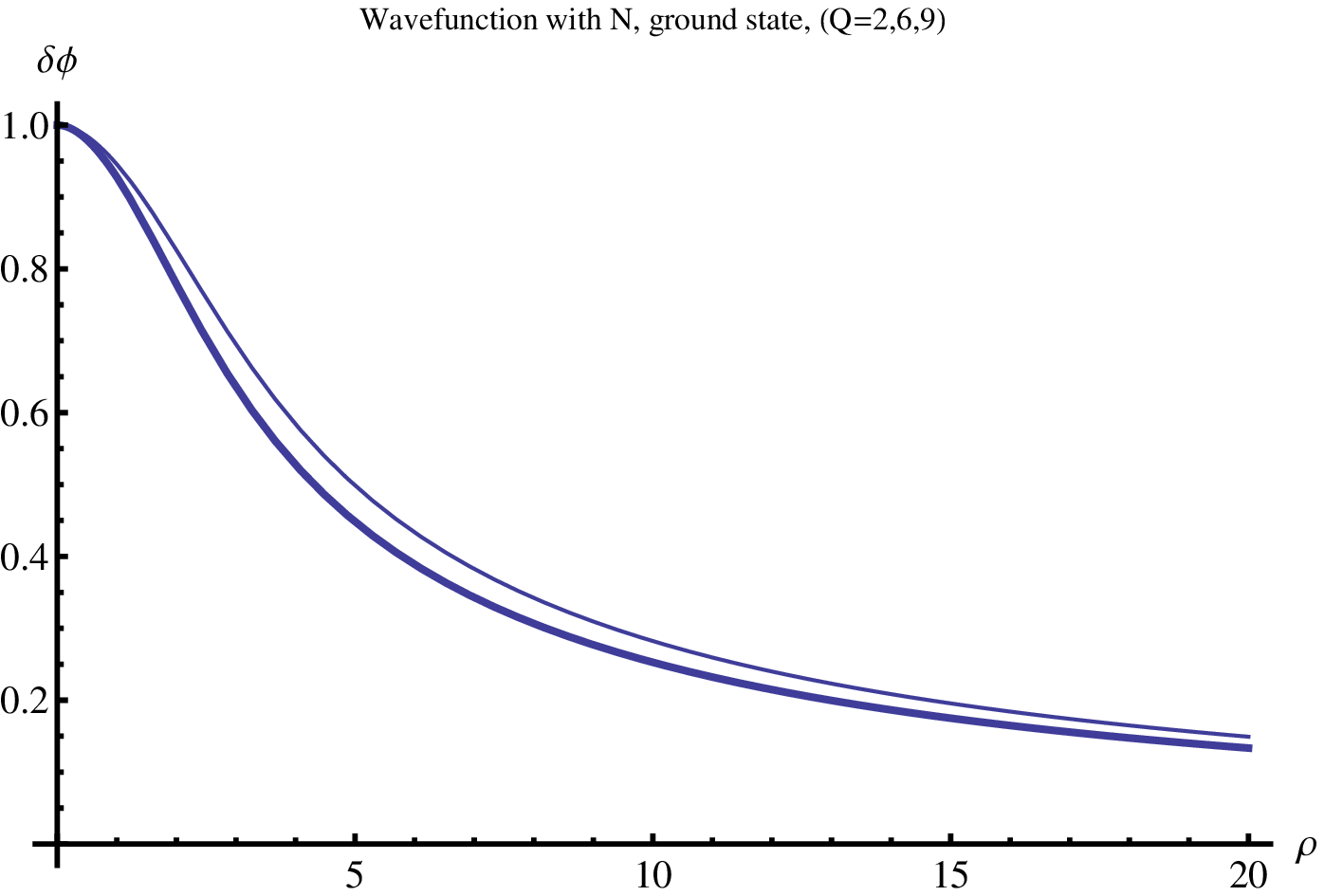}
  \caption{Top : The wave function of $\delta \phi$with Dirichlet (Left) boundary condition and with Neumann (Right) boundary condition. Bottom : The wave function of $\delta \phi$. \label{phiWavefunction}}
\end{center}
\end{figure}

The spectrum is the same as with the Neumann boundary condition at $\tilde{Q}$=0 but different at finite $\tilde{Q}$. The ground-state and the excitation spectra are presented in Fig. \ref{mphimqwithDN1} with two different boundary conditions and $y_v(\infty)$ =0.1 (left), 1(right). The mass computed with D increases when the density is increased, while the mass with N decreases. The origin of this difference can be seen Fig.~\ref{phiWavefunction}. With N, the wave function is more flattened for larger density, so the energy cost is lowered. But with D, the wave function bends increasingly more near the origin when Q becomes large. With D, the system costs more energy to fluctuate from the vacuum and with N, the energy cost is lowered as the density is increased.


\newpage


\end{document}